\documentclass[]{jfm}

\usepackage{amsmath}
\usepackage{amssymb}
\usepackage{psfig}
\usepackage{fancyhdr}
\usepackage{rotate}
\usepackage{rotating}
\usepackage{dcolumn}
\usepackage{bm}
\usepackage{subfigure}
\usepackage{epsfig}
\usepackage{natbib}
\usepackage{graphicx}
\usepackage{color}
\usepackage{latexsym}
\usepackage{amsmath}

\newcommand{\prat}{\nu}
\newcommand{\ymod}{{\cal E}}
\newcommand{\vect}[1]{\mbox{\boldmath$#1$\unboldmath}}

\newcommand{\etal}{{\it et al.}}
\newcommand{\ie}{{\it i.e.}}
\newcommand{\eg}{{\it e.g.}}
\newcommand{\etc}{{\it etc.}}

\newcommand{\bsl}{{\ell}}

\newcommand{\Pe}{\mbox{Pe}}

\newcommand{\kb}{k}
\newcommand{\ka}{\kappa a}

\newcommand{\grad}{\vect{\nabla}}
\newcommand{\lapl}{\nabla^2}
\newcommand{\dive}{\vect{\nabla} \cdot}
\newcommand{\curl}{\vect{\nabla} \times} 

  \title[Displacement of a charged sphere in an elastic Brinkman
    medium]{Electric-field-induced displacement of a charged spherical
    colloid embedded in an elastic Brinkman medium}

\author[]{R\ls E\ls G\ls H\ls A\ls N\ns J.\ns H\ls I\ls L\ls L}
  
  \author[R.~J.~Hill and
    M.~Ostoja-Starzewski]{REGHAN~J.~HILL$^1$\footnote{The author to
    whom correspondence should be addressed. Tel.: +1 514 398-6897,
    Fax: +1 514 398-6678, E-mail: reghan.hill@mcgill.ca} and
    M.~OSTOJA-STARZEWSKI$^2$}
  
  \affiliation{$^1$Department of Chemical Engineering and McGill
    Institute for Advanced Materials, McGill University, Montreal,
    Quebec, H3A 2B2, CANADA $^2$Department of Mechanical Science and
    Engineering, University of Illinois at Urbana-Champaign, Urbana,
    IL 61801, U.S.A.}

\begin{document}
  
  \maketitle


  \begin{abstract}
    When an electric field is applied to an electrolyte-saturated
    polymer gel embedded with charged colloidal particles, the force
    that must be exerted by the hydrogel on each particle reflects a
    delicate balance of electrical, hydrodynamic and elastic
    stresses. This paper examines the displacement of a single charged
    spherical inclusion embedded in an uncharged hydrogel. We present
    numerically exact solutions of coupled electrokinetic transport
    and elastic-deformation equations, where the gel is treated as an
    incompressible, elastic Brinkman medium. This model problem
    demonstrates how the displacement depends on the particle size and
    charge, the electrolyte ionic strength, and Young's modulus of the
    polymer skeleton. The numerics are verified, in part, with an
    analytical (boundary-layer) theory valid when the Debye length is
    much smaller than the particle radius. Further, we identify a
    close connection between the displacement when a colloid is
    immobilized in a gel and its velocity when dispersed in a
    Newtonian electrolyte. Finally, we describe an experiment where
    nanometer-scale displacements might be accurately measured using
    back-focal-plane interferometry. The purpose of such an experiment
    is to probe physicochemical {\em and} rheological characteristics
    of hydrogel composites, possibly during gelation.
  \end{abstract}
  
\section{Introduction}
  
Hydrogels are soft, water-saturated networks of polymer with
molecular-scale porosity. They find widespread use in
molecular-separation technologies (gel electrophoresis), drug
delivery, scaffolds for tissue engineering, cell culture, wound
care~\citep[\eg,][]{Lowman:2004,Yui:2004}, and microfluidic pumping
and control~\citep[\eg,][]{Bassetti:2005}.

This paper concerns a novel class of hydrogel composites where charged
colloidal inclusions are immobilized in an uncharged hydrogel
matrix. \cite{Matos:2006} recently demonstrated that polyacrylamide
hydrogels doped with silica nanoparticles significantly enhance the
electric-field-induced transport of both charged and uncharged
molecules through the composite. The underlying electroosmotic flow
mechanism is consistent with theoretical expectations for weak
perturbations to equilibrium~\citep{Hill:2006b}.

Each charged inclusion experiences electrical, hydrodynamic and
mechanical contact forces, while the mobile counter charge produces an
electro-osmotic flow that permeates the surrounding
polymer. \cite{Hill:2006b} recently developed an electrokinetic
transport model that quantifies how the charge and size of the
inclusions, the ionic strength of the electrolyte, and gel
permeability influence the electro-osmotic pumping capacity of an
uncharged polymer network with randomly dispersed, impenetrable
inclusions. To a first approximation, the elastic distortion of the
network does not influence the flow or ion fluxes, so little is known
of the particle displacements and flow-induced distortion of the
polymer.

The particle displacement could be used as a diagnostic to probe the
physicochemical characteristics of the particle-polymer interface,
much like the electric-field-induced particle velocity
(electrophoretic mobility) is used to infer the so-called
$\zeta$-potential of colloidal particles dispersed in Newtonian
electrolytes. Also, knowledge of the micro-scale strain field in the
hydrogel is essential for establishing the electric field strength
required to initiate micro-scale fracture. Finally, the relationship
between particle displacement and the elastic and viscous
characteristics of the surrounding matrix is central to the rapidly
advancing field of micro-rheology, which seeks to quantify the
dynamics and structure of complex
fluids~\citep{Solomon:2001,Willenbacher:2007}. In this work, our
principal objective is to quantify how the particle, electrolyte and
hydrogel characteristics influence the electric-field-induced particle
displacement.

As a first step, we solve a model problem where classical
electrokinetic transport processes are coupled to the deformation of
an incompressible isotropic, homogeneous porous medium. The analysis
is restricted to situations where the applied electric field is
uniform and weak, so the particle displacement is small and
perturbations to equilibrium may be linearized accordingly. In this
manner, our approximations are similar to those widely adopted in the
classical theories of micro-electrophoresis~\citep{Russel:1989} and
other phoretic motion~\citep{Anderson:1989}. Nevertheless, our
calculations are not restricted by the magnitude of the particle
$\zeta$-potential or the thickness of the equilibrium diffuse layer of
counterions (Debye length).

One promising approach to measure small particle displacements induced
by an electric field involves collecting light scattered from a
micron-sized inclusion at the focus of an optical
trap. Back-focal-plane
interferometry~\citep[\eg,][]{Gittes:1997,Schnurr:1997,Allersma:1998}
provides nanometer-resolution position detection with $1$--$10^4$~Hz
temporal resolution. The technique has been successfully applied to
measure $0.1$--$100$~nm displacements of colloidal inclusions in
complex fluids (\eg, polymer solutions and colloidal dispersions) and
hydrogels, as well as the electric-field-induced velocity
(electrophoretic mobility) of colloidal spheres dispersed in Newtonian
electrolytes~\citep[\eg,][]{Galneder:2001}. Reviews by
\cite{MacKintosh:1999} and \cite{Furst:2005} present authoritative
perspectives of the motivations, development and achievements of
micro-rheology, with emphasis on single-particle dynamics and the use
of optical traps.

The elastic restoring force of stiff gels significantly attenuates
Brownian fluctuations in the position of colloidal particulates
trapped in the hydrogel skeleton. Accordingly, passive
micro-rheological techniques are well suited, and indeed limited, to
weak gels. \cite{Schnurr:1997} showed that the mean-squared
displacement of a particle with radius $a$ trapped in an
incompressible elastic matrix with Young's modulus $\ymod$ is $\kb T /
(2 \pi a \ymod)$ ($\kb T$ is the thermal energy). With $a = 500$~nm
and $\ymod = 10$~kPa, for example, the mean-squared displacement is
$\approx 0.4$~nm. Therefore, to register larger displacements in
stiffer matrices, active micro-rheology typically adopts magnetic or
optical forces to promote oscillatory particle motion. For example,
\cite{Yamaguchi:2005} recently demonstrated the use of a lock-in
amplifier to correlate the particle response in an harmonic optical
trap to resolve extraordinarily small displacements over a wide
frequency range. This proved successful for following the
transformation (gelation) of viscous polymer solutions to moderately
stiff visco-elastic hydrogels.

\cite{Galneder:2001} demonstrated the use of electrical forces to
facilitate single-particle micro-electrophoresis studies. Their
experiments produced $\sim 1$~$\mu$m~s$^{-1}$ particle velocities in
Newtonian electrolytes at $\sim 10$--$10^2$~Hz. Evidently, $\sim
10$--$100$~nm particle displacements were sufficient to measure the
electrophoretic mobility. In principle, however, smaller
electric-field-induced displacements could be resolved by correlating
the response (particle displacement) with the applied electric field.

Accordingly, we complement our principal theoretical results with a
description of a novel experiment that correlates the forcing and
response time series when an electric field is used to impart
quasi-steady oscillatory motion of a charged sphere embedded in a
hydrogel. Because the electric-field-induced displacement reflects the
size and charge of the inclusions, and the ionic strength of the
electrolyte, this experiment could probe physicochemical changes at
the inclusion-hydrogel interface. These are not registered using
conventional laser-tweezer
micro-rheology~\citep[\eg,][]{Valentine:1996,Yamaguchi:2005} because
the optical force of the trap, hydrodynamic drag of the electrolyte,
and the elastic restoring force of an (uncharged) gel are (to a
reasonable first approximation) independent of the inclusion surface
charge~\citep{Gittes:1997,Levine:2001b}.

The theoretical interpretation of such experiments may be complicated
by several practical considerations and a variety of interesting
physical phenomena. For example, if large electric fields are required
to produce measurable particle displacements, then non-linear
electro-kinetic influences, such as electroosmosis of the second
kind~\citep[\eg,][]{Mishchuk:1995} and induced-charge
electroosmosis~\citep[\eg,][]{Bazant:2004} come into play. Aided by
experiments, \cite{Mishchuk:1995} demonstrated that electroosmosis of
the second kind is significant when the applied (uniform) electric
field $E > 25 \kb T / (ae)$ ($e$ is the fundamental charge, so $\kb T
/ e \approx 25$~mV at room temperature). It follows that a linearized
electrokinetic model, such as the one pursued in this paper, should be
reasonable for particles with radius $a = 500$~nm when $E \lesssim
13$~kV~cm$^{-1}$. Depending on the electrolyte conductivity, Joule
heating may become problematic, so lower field strengths are obviously
desirable.

Experiments may also be complicated by electrochemical reactions at
the electrodes. These influences can be minimized by adopting
time-dependent (oscillatory) electric fields, as exploited in
electroacoustic~\citep[\eg,][]{OBrien:1988} and dielectric
relaxation~\citep[\eg,][]{Delacey:1981} diagnostic experiments for
colloidal dispersions, and electrically guided colloidal assembly
processes~\citep[\eg,][]{Ristenpart:2007}. Extending the present
quasi-steady theory to handle oscillatory electric fields at
frequencies where dynamics are important is a significant
challenge. For example, several disparate length scales present an
extremely `stiff' computational problem, and dissipative wave-like
dynamics are expected from the coupling of viscous (electrolyte) and
elastic (polymer) phases.

Other interesting complications, which are not unique to the proposed
electrically forced micro-rheology experiments, arise from polymer
inhomogeneity~\citep[\eg,][]{Maggs:1998,Huh:2006} and the degree of
sticking and slipping at the particle-hydrogel interface. In
principle, these influences can be integrated into the present
computational methodology with appropriate modifications of the
boundary conditions at the particle-polymer interface, and by allowing
the elastic modulus and hydrodynamic permeability of the polymer
skeleton to vary with radial
position~\citep[\eg,][]{Levine:2001b,Chen:2003}. At present, however,
these complications are of secondary importance to the task of
understanding electrokinetic influences with homogeneous polymer.

Finally, our analysis concerns uncharged polymer networks. While many
hydrogel skeletons, especially those of biological origin, bear fixed
charge, poly(acrylamide) and poly(ethylene oxide) hydrogels, for
example, as well as physical gels from block copolymers, are, at least
in principle, uncharged~\citep{Lowman:2004,Yui:2004}. Note that
electric-field-induced actuation of hydrogels is limited exclusively
to poly-electrolyte (intrinsically charged) hydrogels, which often
respond to relatively weak electric fields ($E \sim
10$~V~cm$^{-1}$)~\citep{Shiga:1997}. Although our primary concern is
with micro-rheology, our theory may contribute to understanding how
micro-scale transport processes influence macro-scale deflection
(electrical actuation) of uncharged hydrogel skeletons that are
endowed with charge from particulate inclusions.

Before presenting the full model and examining the results, it is
instructive to first consider the expected displacement $Z$ of an
inclusion if the bare electrical force $\sigma 4 \pi a^2 E$ is
balanced by an elastic restoring force $2 \pi a \ymod Z$. Here,
$\sigma$ is the surface charge density, $4 \pi a^2$ is the surface
area, $E$ is the electric-field strength, and $\ymod$ is Young's
modulus of the gel. Accordingly, the particle displacement
is\footnote{Here, the elastic restoring force is the value when a
rigid sphere with radius $a$ is embedded in an elastic continuum with
Young's modulus $\ymod$ and Poisson's ratio $\prat = 1/2$.}
\begin{equation} \label{eqn:smoluchowski}
  Z = 2 \sigma a \ymod^{-1} E.
\end{equation}
Equation~(\ref{eqn:smoluchowski}) over-estimates the displacement by a
factor of $(2/3) \ka$ when $\ka \rightarrow \infty$ and $|\zeta| < \kb
T / e$. Note that $\ka$ is the particle radius $a$ scaled with the
Debye length $\kappa^{-1}$, $\zeta$ is the $\zeta$-potential, $\kb T$
is the thermal energy, and $e$ is the fundamental charge. When $\ka
\rightarrow 0$ and $|\zeta| < \kb T / e$, however,
Eqn.~(\ref{eqn:smoluchowski}) becomes
\begin{equation} \label{eqn:huckel}
  Z = 2 \zeta \epsilon_s \epsilon_o \ymod^{-1} E \ \ (\ka
  \rightarrow 0, |\zeta| < \kb T / e).
\end{equation}
Equation~(\ref{eqn:huckel}) and the correct form of
Eqn.~(\ref{eqn:smoluchowski}) for $\ka \rightarrow \infty$ (see
section~\ref{sec:results}) are, respectively, reminiscent of the
well-known H{\" u}ckel and Smoluchowski limits for the electrophoretic
mobility~\citep{Russel:1989}. Indeed, despite obvious differences, the
inclusion displacement and electrophoretic velocity have a remarkably
similar and, in general, complicated dependence on the scaled particle
size $\ka$ and scaled $\zeta$-potential $\zeta e / (\kb T)$.

\section{Models for electrokinetic transport and elastic deformation}

\begin{figure}
  \begin{center}
    \input{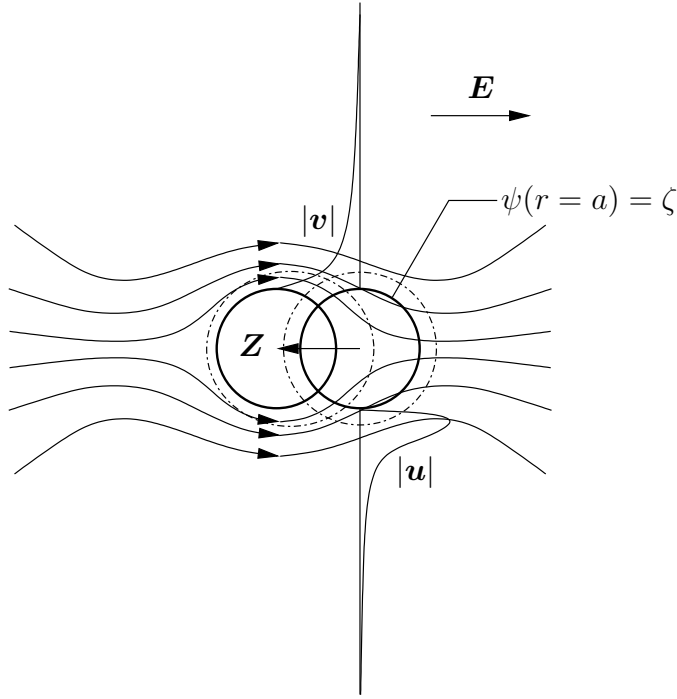}
  \end{center}
  \caption{\label{fig:schematic} Schematic representation of a
    (negatively) charged, spherical colloid embedded in an unbounded
    electrolyte-saturated, elastic polymer gel (elastic Brinkman
    medium). An electric field $\vect{E}$ is applied that drives
    electro-osmotic flow $\vect{u}$, which, in turn, exerts a body
    force on the gel that induces a displacement field $\vect{v}$. The
    net result of the electrical, hydrodynamic and elastic stress is a
    displacement $\vect{Z}$ of the particle (to the left). The dashed
    lines denote the equilibrium (right) and polarized (left) diffuse
    double layers. In the far-field ($r \rightarrow \infty$), the
    velocity and displacement fields decay as $r^{-3}$. In this work,
    the displacement field is treated as the sum of two fields that
    each decay as $r^{-1}$; one is induced by a particle displacement
    $\vect{Z} \ne \vect{0}$ with $\vect{E} = \vect{0}$, and the other
    is the flow-induced distortion with $\vect{E} \ne \vect{0}$ and
    $\vect{Z} = \vect{0}$.}
\end{figure}

As depicted in figure~\ref{fig:schematic}, we consider a single
spherical colloid with radius $a$ and surface charge density $\sigma$
embedded in an unbounded, electrically neutral hydrogel. The gel is
modeled as a homogeneous Brinkman medium that is saturated with an
aqueous electrolyte (\eg, NaCl). Together, the electrolyte
concentration, surface charge density and particle radius manifest as
an electrostatic potential $\zeta$ at the colloid surface ($r =
a$). Note that the counter-charge is concentrated in a diffuse layer
with thickness (Debye length) $\kappa^{-1}$.

\subsection{Electrokinetic transport}

The electrokinetic model used in this work to calculate the
equilibrium and perturbed electrolyte ion concentrations, pressure,
and fluid velocity is a straightforward extension of the {\em standard
electrokinetic model}~\citep{Booth:1950} widely used to describe
micro-electrophoresis and other electrokinetic phenomena. Details have
been presented elsewhere~\citep{Hill:2006b,Hill:2006a}. The full set
of (steady) transport equations is
\begin{eqnarray} \label{eqn:pbeqn}
  \epsilon_o \epsilon_s \lapl \psi = - \sum_{j=1}^{N} n_j z_j e \\
  \vect{j}_j = n_j \vect{u} - z_j e n_j \frac{D_j}{\kb T} \grad \psi -
  D_j \grad{n_j}\\ \eta \lapl \vect{u} - \grad p = (\eta / \bsl^2)
  \vect{u} + \sum_{j=1}^{N} n_j z_j e \grad \psi, \label{eqn:nseqn}
\end{eqnarray}
with (steady) ion and electrolyte conservation equations
\begin{equation}
  \dive{\vect{j}_j} = 0 \mbox{ and } \dive{\vect{u}} = 0. \nonumber
\end{equation}
The electrostatic potential, $N$ ion concentrations and fluxes,
electrolyte velocity, and pressure are denoted $\psi$, $n_j$ and
$\vect{j}_j$, $\vect{u}$, and $p$, respectively. Other variables are
the solvent dielectric constant $\epsilon_s$, permitivity of a vacuum
$\epsilon_o$, fundamental charge $e$, ion valances $z_j$ and diffusion
coefficients $D_j$, thermal energy $\kb T$, solvent viscosity $\eta$,
and Darcy permeability $\bsl^2$ (square of the \cite{Brinkman:1947}
screening length). Note that the double-layer thickness (Debye length)
is
\begin{equation}
  \kappa^{-1} = \sqrt{\kb T \epsilon_s \epsilon_o / (2 I e^2)},
  \nonumber
\end{equation}
where $I = (1/2) \sum_{j=1}^{N} z^2_j n^\infty_j$ is the bulk ionic
strength, with $n_j^\infty$ the bulk ion concentrations. Since this
work deals exclusively with steady (or quasi-steady) flows, the fluid
velocity in Eqn.~(\ref{eqn:nseqn}) is relative to a stationary polymer
skeleton.

The equations are solved by perturbing (to linear order) from an
equilibrium base state where $\vect{u} = \vect{0}$ and the equilibrium
electrostatic potential, ion concentrations and pressure are denoted
$\psi^0$, $n_j^0$, and $p^0$,
respectively~\citep{OBrien:1978,Hill:2003a}. With the application of a
uniform electric field $\vect{E}$,
\begin{equation}
  \psi = \psi^0 + \psi', \ n_j = n_j^0 + n_j',
  \ \vect{u} = \vect{u}', \nonumber
\end{equation}
where
\begin{equation}
  \psi' = - \vect{E} \cdot \vect{r} + \hat{\psi}(r) \vect{E} \cdot
  \vect{e}_r, \ n_j' = \hat{n}_j(r) \vect{E} \cdot \vect{e}_r,
  \nonumber
\end{equation}
and
\begin{eqnarray}
  \vect{u} &=& \grad \times \grad \times f(r) \vect{E} \nonumber \\
  &=& - 2 f_r r^{-1} (\vect{E} \cdot \vect{e}_r) \vect{e}_r - (f_{rr}
  + f_{r} r^{-1}) (\vect{E} \cdot \vect{e}_\theta)
  \vect{e}_\theta. \label{eqn:pert1}
\end{eqnarray}
Here, $f_r = \mbox{d} f / \mbox{d}r$, $f_{rr} = \mbox{d}^2 f /
\mbox{d}r^2$, $\vect{r}$ is position in a spherical polar coordinate
system $(r, \theta, \phi)$ with unit basis vectors ($\vect{e}_r,
\vect{e}_\theta, \vect{e}_\phi)$, and $\vect{e}_z$ is the orientation
of the polar axis (so $\vect{e}_r \cdot \vect{e}_z = \cos{\theta}$).

At the particle-electrolyte interface ($r = a$),
\begin{equation}
  \vect{u} = \vect{0} \ (\mbox{no slip}), \ \vect{j}_j \cdot
  \vect{e}_r = 0 \ (\mbox{no flux}), \nonumber
\end{equation}
and
\begin{equation}
  \epsilon_s \epsilon_o \grad_>{\psi} \cdot \vect{e}_r - \epsilon_p
  \epsilon_o \grad_<{\psi} \cdot \vect{e}_r = -\sigma \
  (\mbox{constant surface charge}), \nonumber
\end{equation}
where $\epsilon_p$ is the particle dielectric constant, and $\sigma$
is the (constant) surface charge density; the subscripts attached to
the gradient operators distinguish the particle ($<$) and solvent
($>$) sides of the interface.

In the far-field ($r \rightarrow \infty$),
\begin{eqnarray} \label{eqn:farfield1}
  \psi &\rightarrow& - \vect{E} \cdot \vect{r} + (\vect{E} \cdot
  \vect{e}_r) D^E r^{-2} \nonumber \\ n_j &\rightarrow& n_j^\infty +
  (\vect{E} \cdot \vect{e}_r) C^E_j r^{-2} \nonumber \\ \vect{u}
  &\rightarrow& - 2 C^E r^{-3} (\vect{E} \cdot \vect{e}_r) \vect{e}_r
  - C^E r^{-3} (\vect{E} \cdot \vect{e}_\theta) \vect{e}_\theta,
\end{eqnarray}
where the scalar coefficients $D^E$, $C_j^E$ and $C^E \eta / \bsl^2$
are, respectively, the dipole strengths of the electrostatic
potential, ion concentrations, and pressure perturbations, induced by
the electric field. Note that the far-field flow is a Darcy flow,
$\vect{u} = - (\bsl^2 / \eta) \grad{p'}$, decaying as $r^{-3}$ as $r
\rightarrow \infty$.

The full equations and boundary conditions above are the basis of an
efficient numerical solution that yields the so-called asymptotic
coefficients $D^E$, $C_j^E$ and $C^E$, which have been used to
calculate the bulk {\em electrical conductivity} and {\em pore
mobility} of dilute polymer-gel
composites~\citep{Hill:2006b,Hill:2006c}. This work draws upon $C^E$,
and introduces a new asymptotic coefficient $Z^E$ to characterize the
far-field decay of the electric-field-induced elastic displacement
field.

\subsection{Elastic deformation}

Elastic deformation of the polymer gel is calculated by modeling the
polymer skeleton as an elastic Brinkman medium, whose equation of
static equilibrium is
\begin{equation} \label{eqn:eqm}
  \dive{\vect{\sigma}} + (\eta / \bsl^2) \vect{u} = \vect{0}.
\end{equation}
Here, the elastic stress tensor is~\citep{Landau:1986}
\begin{equation} \label{eqn:elasticstress}
  \vect{\sigma} = \frac{\ymod}{(1 + \prat)} [\vect{e} +
    \frac{\prat}{(1 - 2 \prat)} (\dive{\vect{v}})
    \vect{\delta}],
\end{equation}
where $\vect{v}$ is the (small-amplitude) displacement, $\ymod$ is
Young's modulus, $\prat$ is Poisson's ratio, $\vect{e} = (1/2)
[\grad{\vect{v}} + (\grad{\vect{v}})^T]$, and $\vect{\delta}$ is the
identity tensor. Substituting Eqn.~(\ref{eqn:elasticstress}) into
Eqn.~(\ref{eqn:eqm}) gives
\begin{equation} \label{eqn:staticeqm}
 \frac{\ymod}{2 (1 + \prat)} \lapl{\vect{v}} + 
\frac{\ymod}{2 (1 + \prat) (1 - 2 \prat)} \grad{(\dive{\vect{v}})} =
 - (\eta / \bsl^2) \vect{u}. 
\end{equation}
Again, the fluid velocity is relative to a stationary polymer skeleton
($\partial \vect{v} / \partial t = \vect{0}$).

When the (leading order) displacement is divergence-free, which is the
situation addressed throughout this paper, deformation can only
influence the (isotropic) permeability tensor $(\eta / \bsl^2)
\vect{\delta}$ by inducing anisotropy. Note that any
deformation-induced change in permeability yields a non-zero product
of the permeability perturbation and the fluid velocity, the later of
which is itself a perturbation. Accordingly, these second-order terms
are neglected in the present (linearized) theory.

There is also a possibility of anisotropy in permeability due to the
underlying random microstructure of the medium (hydrogel matrix),
which itself is not isotropic at very small length scales. However, we
assume this medium to be statistically isotropic and homogeneous, so
that, at the level of a Representative Volume Element (RVE) of the
deterministic continuum, the anisotropy vanishes just as the
fluctuations in constitutive response tend to zero;
see~\cite{Ostoja-Starzewski:1989} for a random elastic model.  Such a
scale-dependent homogenization (\ie, a passage from a random
microstructure to the RVE) was recently studied in the context of
Stokesian permeability~\citep{Du:2006}, albeit the departure from
anisotropy was not addressed explicitly; see
also~\cite{Ostoja-Starzewski:2007} for related studies in many other
material problems.

The Poisson ratios of several widely used, highly swollen, transparent
hydrogels are reported greater than 0.45. In particular, poly(vinyl
alcohol) hydrogels prepared with a mixed solvent of dimethyl sulfoxide
and water have $\prat \approx 0.472$~\citep{Urayama:1993}, and
polyacrylamide hydrogels have $\prat \approx
0.457$~\citep{Takigawa:1996}. However, it is important to note that
these measurements are ascertained from macroscale experiments where
the characteristic length and time scales cannot probe the equilibrium
(long-time) state of strain. For example, the fractional change in
volume after relaxing to equilibrium is $\delta V / L^3 \sim (1-2
\prat) l / L$, where the strain $l / L \ll 1$ is the ratio of the
imposed displacement $l$ to the specimen size $L$. The flux of solvent
flowing through the specimen during the so-called draining time $\tau$
is $u_c \sim \delta V / (L^2 \tau) \sim (1-2 \prat) l /
\tau$. Further, the flux is driven by a pressure gradient $p_c / L
\sim (\eta / \bsl^2) u_c$, where, to balance the elastic stresses, the
characteristic pressure is $p_c \sim \ymod (l / L)$. Together, the
foregoing yield
\begin{equation} \label{eqn:draining}
  \tau \sim (1-2\prat) (\eta / \ymod) (L/\bsl)^2,
\end{equation}
so with $\eta \sim 10^{-3}$~Pas, $\ymod \sim 10^5$~Pa, $L \sim
10^{-2}$~m (macro-scale experiment) and $\bsl \sim 10^{-9}$~m, the
relaxation time is $\tau \sim (1 - 2 \prat) 10^6$~s. Clearly, this is
extraordinarily long if $\prat$ is not sufficiently close to
$0.5$. Accordingly, when the experimental time scale (\eg, reciprocal
frequency) is shorter than the draining time, the change in volume
will be smaller than at equilibrium, and the apparent Poisson ratio
will be greater than the drained value.

In contrast, the draining time associated with the displacement of a
microsphere ($L = a \sim 10^{-6}$~m) embedded in a hydrogel with
$\ymod \sim 10^5$~Pa and $\bsl \sim 10^{-9}$~m is only $\tau \sim (1 -
2 \prat) 10^{-2}$~s. Clearly, such an experiment is much better suited
to probing the compressibility of the polymer skeleton. However,
solving the problem with a compressible polymer matrix demands a
distinctly different computational methodology than the one adopted in
this paper for incompressible gels. Moreover, compressibility is
anticipated to yield {\em quantitative}---not qualitative---changes in
the calculated particle displacement. This expectation is supported,
in part, by the fact that, in the absence of an electric field, the
quasi-steady particle displacement varies by at most 25 percent from
the incompressible limit as Poisson's ratio spans the range
$0$--$0.5$\footnote{Assuming a constant {\em shear modulus}
$\ymod/[2(1+\prat)]$~\citep{Landau:1986}.}~\citep{Schnurr:1997}. Clearly,
a definitive answer requires a solution of the electrokinetic
equations with arbitrary Poisson's ratio: an ambitious goal that is
beyond the relatively modest scope of the present work.

Finally, the density and, hence, rigidity of the inclusions (\eg,
polymer latex or silica) are typically much greater than those of the
gel. Accordingly, the inclusions are treated as rigid
spheres. Moreover, the displacement field is assumed to be continuous
across the inclusion-hydrogel interface. Other interfacial conditions
include the possibility of (tangential) slip~\citep[\eg][]{Mura:1985}
or an opening crack, for example. The opening of a crack significantly
complicates the analysis, and a slipping boundary condition is
difficult to justify here, since it requires a physical mechanism to
exert an interfacial radial stress while maintaining zero (relative)
radial displacement and zero tangential stress. Accordingly, neither
possibility is pursued here.

\section{Superposition to calculate the particle displacement}

It is convenient to calculate the electric-field-induced particle
displacement $\vect{Z}$ by superposing two linearly independent
displacement fields.

One is the displacement field induced by a small displacement
$\vect{Z}$ of the inclusion in the absence of an electric field
($\vect{E} = \vect{0}$). There is no Darcy drag, and the solution of
Eqn.~(\ref{eqn:staticeqm}) can be calculated analytically (see
appendix~\ref{app:1}). The resulting mechanical-contact force exerted
by the polymer {\em on} the particle is\footnote{With a slipping
boundary condition (\ie, zero radial displacement and zero tangential
stress at $r = a$), the force is $-12 \pi \ymod a \vect{Z} (1 -
\prat) / [(7 - 8 \prat)(1 + \prat)]$.}
\begin{equation} \label{eqn:force1}
  \vect{f}^{m,Z} = - \frac{2 \pi a \ymod \vect{Z} (1-\prat)}{(5/6 -
    \prat)(1 + \prat)}.
\end{equation}

The other arises from the Darcy drag force when an electric field
$\vect{E}$ is applied and the inclusion is fixed at the origin
($\vect{Z} = \vect{0}$). The Darcy drag force in
Eqn.~(\ref{eqn:staticeqm}) is calculated from the electrokinetic
transport equations with a rigid polymer gel. Then the displacement
can be obtained by solving Eqn.~(\ref{eqn:staticeqm}). This is
detailed in section~\ref{sec:force}, where it is also shown that the
mechanical-contact force exerted by the polymer {\em on} the inclusion
is
\begin{equation} \label{eqn:force2}
  \vect{f}^{m,E} =(8/3) \pi Z^E \ymod \vect{E} - 4 \pi (\eta /
  \bsl^2) C^E \vect{E} \ \ (\prat = 1/2).
\end{equation}
In addition to the net mechanical-contact force
\begin{equation}
  \vect{f}^m = \vect{f}^{m,Z} + \vect{f}^{m,E},
\end{equation}
there are electrical and hydrodynamic (drag) forces acting on the
particle, denoted $\vect{f}^{e,E}$ and $\vect{f}^{d,E}$,
respectively. These are already known from earlier solutions of the
electrokinetic transport equations with a rigid (unperturbed) polymer
gel~\citep{Hill:2006a,Hill:2006b}. Their sum can be written in terms
of the asymptotic coefficient $C^E$ that characterizes the far-field
decay of the electric-field-induced flow:
\begin{equation} \label{eqn:force3}
  \vect{f}^{e,E} + \vect{f}^{d,E} = 4 \pi (\eta / \bsl^2) C^E
  \vect{E}.
\end{equation}

Finally, static equilibrium of the particle demands
\begin{equation} \label{eqn:fbal}
  \vect{f}^{e,E} + \vect{f}^{d,E} + \vect{f}^{m,Z} + \vect{f}^{m,E} =
  \vect{0}.
\end{equation}
Therefore, collecting the explicit expressions for the various forces
above [Eqns.~(\ref{eqn:force1}), (\ref{eqn:force2}) (with $\prat =
1/2$) and (\ref{eqn:force3})] gives the particle displacement
\begin{equation} \label{eqn:displacement}
  \vect{Z} = (4/3) (Z^E/a) \vect{E} \ \ (\prat = 1/2).
\end{equation}

The task of calculating $Z^E$ is detailed in the next section. Note
that the contributions involving $C^E$ vanish, indicating that the
slowest ($r^{-1}$) far-field decay of the displacement field vanishes
upon superposition.  In other words, there is no net force acting on
the polymer, so the Darcy drag force exerted by the electrolyte on the
polymer is counterbalanced by the mechanical-contact force exerted by
the inclusion on the polymer. In a composite with a finite inclusion
number density, or finite volume fraction $\phi$, part of the net
mechanical-contact force acting on the polymer must be provided by a
mechanical support to balance an accompanying $O(\phi)$ average
pressure gradient~\citep{Hill:2006b}.

\section{The electric-field-induced mechanical-contact force for an incompressible, elastic Brinkman medium} \label{sec:force}

This section addresses the displacement induced by Darcy drag when an
electric field is applied and the inclusion is fixed at the
origin. This problem is adopted to calculate the force
$\vect{f}^{m,E}$ appearing in Eqn.~(\ref{eqn:force2}). Note that the
numerical solution is limited to incompressible ($\dive{\vect{v}} =
0$) displacement fields, so $\prat = 1/2$.

In appendix~\ref{app:elastic}, the displacement field is expanded as a
power series in a small parameter $\epsilon = 1 - 2 \prat$, \ie,
\begin{equation}
  \vect{v} = \vect{v}_0 + \epsilon \vect{v}_1 + ....
\end{equation}
The leading contribution to the displacement $\vect{v}_0$ is
divergence-free ($\dive{\vect{v}_0} = 0$) and satisfies the
$O(\epsilon)$ equation of static equilibrium,
\begin{equation} \label{eqn:oepsilonp}
  \lapl \vect{v}_0 + \grad{(\dive{\vect{v}_1})} = - (\eta / \bsl^2)
  \vect{u} \frac{3}{\ymod} \ \ (\prat = 1/2).
\end{equation}

Incompressibility (as required by the $O(1)$ problem) is guaranteed by
writing
\begin{equation}\label{eqn:edisplacement}
  \vect{v}_0 = \curl{} \curl{} g(r) \vect{E},
\end{equation}
where $g(r)$ is a function of radial position $r$. It follows that
\begin{eqnarray}
  \vect{v}_0 = - 2 g_r r^{-1} (\vect{E} \cdot \vect{e}_r) \vect{e}_r -
  (g_{rr} + g_{r} r^{-1}) (\vect{E} \cdot \vect{e}_\theta)
  \vect{e}_\theta,
\end{eqnarray}
where, for example, $g_r = \mbox{d} g / \mbox{d}r$.

Because the fluid is also incompressible, its velocity field may be
written as
\begin{equation} \label{eqn:fvelocity}
  \vect{u} = \curl{} \curl{} f(r) \vect{E},
\end{equation}
where $f(r)$ is available from earlier work examining the influence of
an electric field~\citep{Hill:2006b} and a bulk concentration
gradient~\citep{Hill:2006a} with a rigid polymer gel ($\vect{v} =
\vect{0}$).

If the deformation is assumed not to affect electrokinetic transport
processes, which is a reasonable approximation when the displacement
is divergence-free (so the polymer segment density and, hence, the
Darcy permeability are unperturbed), then the earlier calculations
also provide an exact solution with (weak) elastic deformation.

Substituting Eqns.~(\ref{eqn:edisplacement}) and~(\ref{eqn:fvelocity})
into the curl of Eqn.~(\ref{eqn:oepsilonp}) gives
\begin{equation}
  \frac{\mbox{d}}{\mbox{d}r} \lapl \lapl g + \frac{3 \eta}{\ymod
  \bsl^2} \frac{\mbox{d}}{\mbox{d}r} \lapl f = 0,
\end{equation}
where
\begin{equation}
  \lapl = \frac{1}{r^2} \frac{\mbox{d}}{\mbox{d}r} (r^2
  \frac{\mbox{d}}{\mbox{d}r}),
\end{equation}
and $f(r)$ is known. In this manner, $\vect{v}_0$ is decoupled from
$\vect{v}_1$.

The fourth-order ordinary differential equation for $g_r(r)$ is solved
numerically as two second-order differential equations\footnote{A
  finite-difference scheme based on Hill~\etal's
  methodology~\citep{Hill:2003a} is adopted. This features an adaptive,
  non-uniform grid to handle the disparate length scales.}:
\begin{eqnarray} \label{eqn:displacementode1}
  \frac{\mbox{d}^2 h}{\mbox{d}r^2} + \frac{4}{r} \frac{\mbox{d}
    h}{\mbox{d}r} - \frac{4}{r^2} h &=& - \frac{3 \eta}{\ymod \bsl^2}
    (\frac{\mbox{d}^2 f_r}{\mbox{d}r^2} + \frac{2}{r} \frac{\mbox{d}
    f_r}{\mbox{d}r} - \frac{2}{r^2} f_r) \\ \frac{\mbox{d}^2
    g_r}{\mbox{d}r^2} &=& h.
\end{eqnarray}

When the displacement is continuous across the inclusion-hydrogel
interface, and vanishes in the far-field, the boundary conditions are
\begin{equation}
  \vect{v}_0 = \vect{0} \mbox{ at } r = a,
\end{equation}
and
\begin{eqnarray}
  \vect{v}_0 \rightarrow - 2 Z^E r^{-1} (\vect{E} \cdot \vect{e}_r)
  \vect{e}_r - Z^E r^{-1} (\vect{E} \cdot \vect{e}_\theta)
  \vect{e}_\theta \mbox{ as } r \rightarrow \infty.
\end{eqnarray}
The asymptotic coefficient $Z^E$ characterizes the strength of the
$r^{-1}$ decay of $\vect{v}$ (reflecting a net force). It follows that
\begin{equation}
  g_r = g_{rr} = 0 \mbox{ at } r = a,
\end{equation}
and
\begin{equation}
  g_r \rightarrow Z^E \mbox{ and } g_{rr} \rightarrow 0 \mbox{ as } r
  \rightarrow \infty.
\end{equation}

An analytical boundary-layer analysis that solves the problem when
$\ka \gg 1$, $\bsl \ll a$ and $|\zeta| < \kb T/ e$ serves to verify
the numerical solution and to highlight the parametric scaling of
$Z$. The result is presented in section~\ref{sec:results}
[Eqn.~(\ref{eqn:bltheory})], where we also examine numerically exact
solutions of the full model.

Turning to the force, the leading contribution to the isotropic stress
requires knowledge of the $O(\epsilon)$ displacement field
$\vect{v}_1$, which is not divergence-free (see
appendix~\ref{app:elastic}). Clearly, the divergence of $\vect{v}_1$
is necessary to evaluate the leading contribution to the
force. Fortunately, the isotropic stress can be obtained from the
solution of the $O(\epsilon)$ problem above. This is achieved by
integrating Eqn.~(\ref{eqn:oepsilon}) once $\vect{v}_0$ is known. The
task is simplified even further, because only the far-field decay of
the displacement and fluid velocity fields are needed to evaluate the
force. We have verified our general procedure by applying it to two
simpler problems: one is the classical problem of Stokes flow past a
sphere, and the other is the elastic restoring force on a rigid sphere
embedded in an incompressible elastic continuum. Recall, the exact
solution of the latter problem, for any $\prat$ [see
Eqn.~(\ref{eqn:force1})], is worked out in appendix~\ref{app:1}.

The mechanical-contact force exerted by the polymer {\em on} the
inclusion is
\begin{equation} \label{eqn:mechforce}
  \vect{f}^{m,E} = \int_{r=a} \vect{\sigma} \cdot \hat{\vect{n}}
  \mbox{d}A =\int_{r \rightarrow \infty} \vect{\sigma} \cdot
  \hat{\vect{n}} \mbox{d}A + \int_{r=a}^{\infty} (\eta / \bsl^2)
  \vect{u} \mbox{d}V.
\end{equation}
Note that (see appendix~\ref{app:elastic})
\begin{equation}
  \vect{\sigma} = \frac{2 \ymod}{3} [\vect{e}_0 + \frac{1}{2}
    (\dive{\vect{v}_1}) \vect{\delta}],
\end{equation}
where $\vect{e}_0 = (1/2) [\grad{\vect{v}_0} +
  (\grad{\vect{v}_0})^T]$, from Eqn.~(\ref{eqn:oepsilonp})
\begin{equation}
  \dive{\vect{v}_1} = -\int_{\infty}^r [(\lapl{\vect{v}_0}) + (3 /
    \ymod) (\eta / \bsl^2) \vect{u}] \cdot \vect{e}_r \mbox{d}r,
\end{equation}
and
\begin{eqnarray}
  \vect{u} &=& - 2 f_r r^{-1} (\vect{E} \cdot \vect{e}_r) \vect{e}_r -
  (f_{rr} + f_{r} r^{-1}) (\vect{E} \cdot \vect{e}_\theta)
  \vect{e}_\theta \nonumber \\ &\rightarrow& - 2 C^E r^{-3} (\vect{E}
  \cdot \vect{e}_r) \vect{e}_r - C^E r^{-3} (\vect{E} \cdot
  \vect{e}_\theta) \vect{e}_\theta \mbox{ as } r \rightarrow \infty.
\end{eqnarray}
Recall, $C^E$ is the asymptotic constant that represents the dipole
strength of the far-field pressure field (decaying as $r^{-2}$) that
drives the far-field Darcy flow (decaying as $r^{-3}$).

Evaluating the first integral on the right-hand side of
Eqn.~(\ref{eqn:mechforce}) over the surface of a large concentric
sphere gives
\begin{equation} \label{eqn:int1}
  \int_{r \rightarrow \infty} \vect{\sigma} \cdot \hat{\vect{n}}
  \mbox{d}A = (8/3) \pi Z^E \ymod \vect{E} - (4/3) \pi (\eta /
  \bsl^2) C^E \vect{E} \ \ (\prat = 1/2).
\end{equation}
The volume integral can be transformed to another integral over the
surface of a large concentric sphere ($\dive{\vect{u}} = 0$ and
$\vect{u}(r=a) = \vect{0}$) giving
\begin{equation} \label{eqn:int2}
  \int_{r=a}^{\infty} (\eta / \bsl^2) \vect{u} \mbox{d}V = (\eta /
  \bsl^2) \int_{r \rightarrow \infty} (\vect{x} \cdot \vect{u})
  \hat{\vect{n}}\mbox{d}A = - (8/3) \pi (\eta / \bsl^2) C^E \vect{E}.
\end{equation}
Finally, adding Eqns.~(\ref{eqn:int1}) and~(\ref{eqn:int2}) gives the
mechanical-contact force as it appears is Eqn.~(\ref{eqn:force2}).

\section{Results} \label{sec:results}

When solving the equations numerically, the characteristic scales
adopted for length, velocity and displacement are
\begin{equation} 
  \kappa^{-1}, \ u^* = \epsilon_s \epsilon_o (\kb T / e)^2 / (\eta a)
  \mbox{ and } \eta u^* / \ymod = \epsilon_s \epsilon_o (\kb T /
  e)^2 / (\ymod a), \nonumber
\end{equation}
respectively. It is therefore convenient to introduce a dimensionless
asymptotic coefficient $\hat{Z}^E$ so
\begin{equation}
  Z^E = \frac{\epsilon_s \epsilon_o (\kb T /e) a}{\ymod (\ka)^2}
  \hat{Z}^E.
\end{equation}
Accordingly, the particle displacement [Eqn.~(\ref{eqn:displacement})]
is
\begin{equation}
  \vect{Z} = (4/3) \frac{\epsilon_s \epsilon_o (\kb T / e)}{\ymod
    (\ka)^2} \hat{Z}^E \vect{E}.
\end{equation}

The independent dimensionless parameters adopted below are $\ka$,
$\zeta e / (\kb T)$ and $\kappa \bsl$. Note that $\hat{Z}^E$ is
independent of $\ymod$, and, furthermore, we will see that the
displacement is a very weak function of $\bsl$. Using a boundary-layer
approximation for $C^E$~\citep{Hill:2006c}, an analytical solution for
$Z^E$, when $\ka \gg 1$, $\bsl \ll a$ and $|\zeta| < \kb T /e$, yields
\begin{equation} \label{eqn:bltheory}
  \vect{Z} \rightarrow 3 \zeta \epsilon_s \epsilon_o \ymod^{-1}
  \vect{E} \ \ \mbox{as} \ \ \ka \rightarrow \infty,
\end{equation}
which is the counterpart to Eqn.~(\ref{eqn:huckel}) identified in the
introduction.

\subsection{Particle displacement with NaCl electrolytes}

\begin{figure}
  \begin{center}
    \vspace{1.0cm}
    \includegraphics[width=6cm]{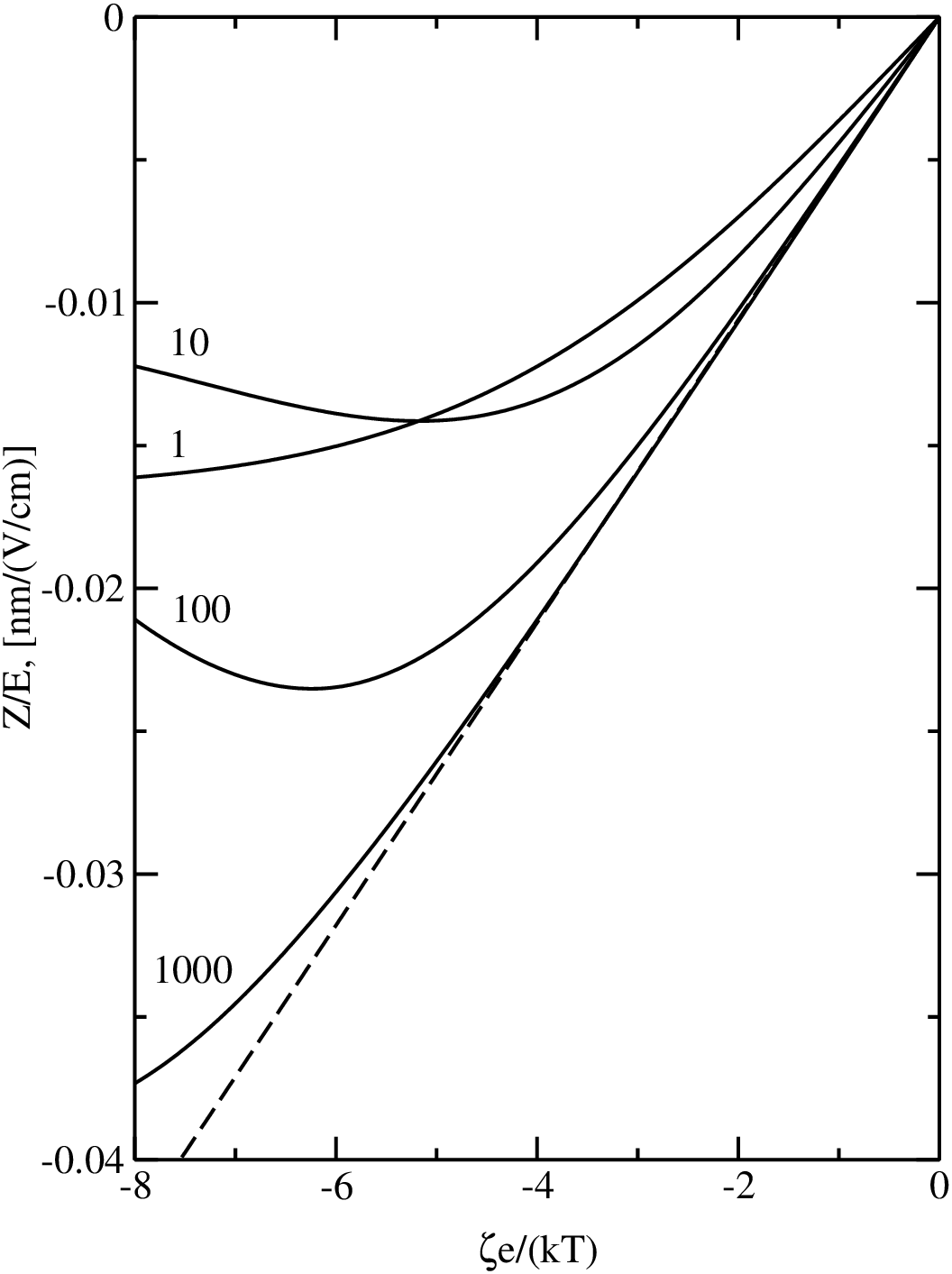}\\
    \vspace{0.5cm}
    \caption{\label{fig:1} The ratio of the displacement $Z$ (nm) to
      the electric field strength $E$ (V~cm$^{-1}$) as a function of the
      scaled $\zeta$-potential $\zeta e / (\kb T)$ for various scaled
      reciprocal double-layer thicknesses $\ka = 1$, 10, 100 and 1000:
      NaCl at $T = 298$~K; $a = 500$~nm; $\bsl = 5$~nm; $\ymod =
      1$~kPa; $\prat = 1/2$. Note that the displacement is inversely
      proportional to Young's modulus $\ymod$. The dashed-line is
      the boundary-layer theory [Eqn.~(\ref{eqn:bltheory})].}
  \end{center}
\end{figure}

\begin{figure}
  \begin{center}
    \vspace{1.0cm}
    \includegraphics[width=6cm]{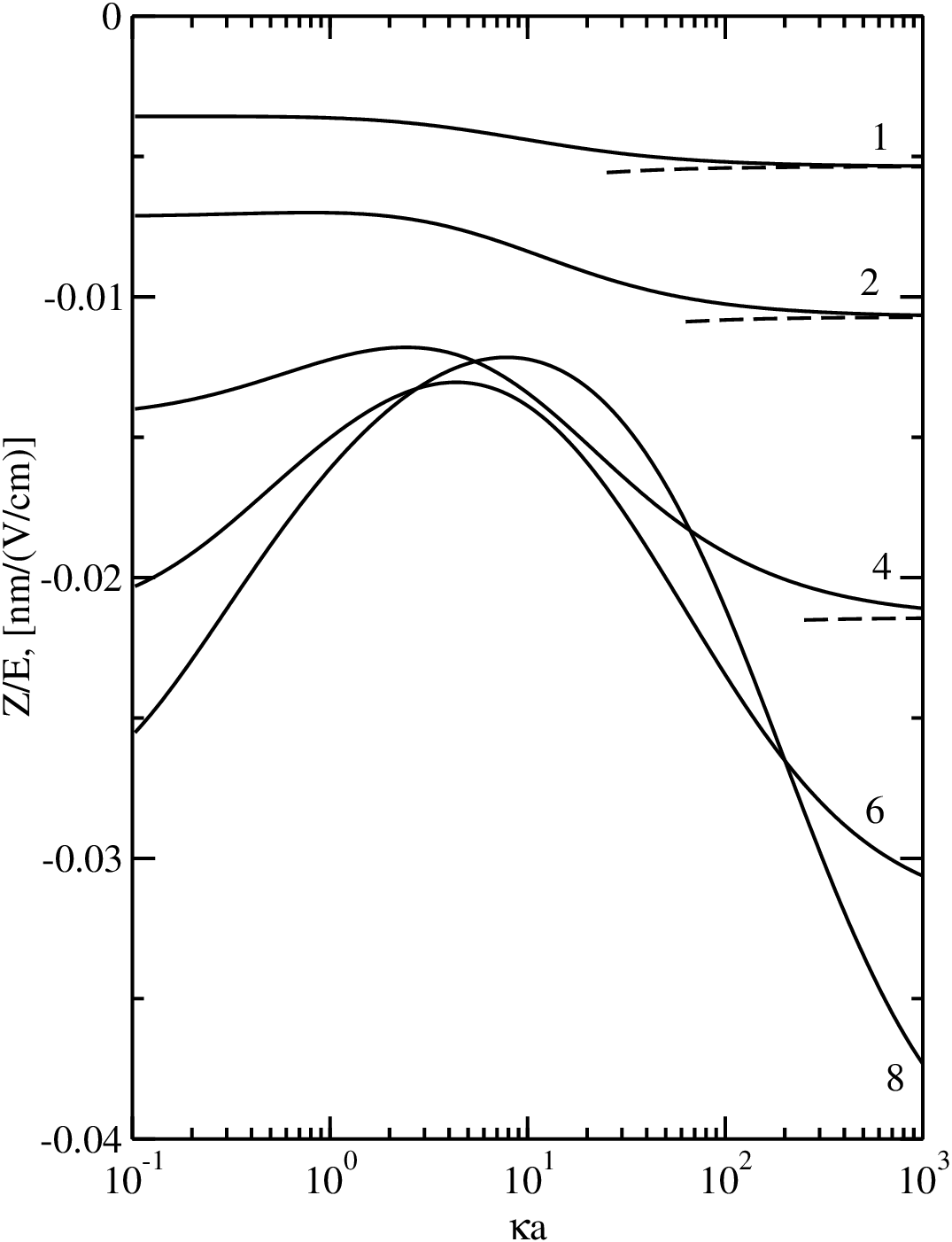}\\
    \vspace{0.5cm}
    \caption{\label{fig:2} The ratio of the displacement $Z$ (nm) to
      the electric field strength $E$ (V~cm$^{-1}$) as a function of the
      scaled reciprocal double-layer thickness $\ka$ for various
      scaled $\zeta$-potentials $-\zeta e / (\kb T) = 1$, 2, 4, 6 and
      8: NaCl at $T = 298$~K; $a = 500$~nm; $\bsl = 5$~nm; $\ymod =
      1$~kPa; $\prat = 1/2$. Note that the displacement is
      inversely proportional to Young's modulus $\ymod$. The
      dashed-lines are the boundary-layer theory
      [Eqn.~(\ref{eqn:bltheory})].}
  \end{center}
\end{figure}

To draw a closer connection to experiments, $Z / E$ is plotted in
figures~\ref{fig:1} and~\ref{fig:2} with $\ymod = 1$~kPa. Therefore,
actual displacements $Z$~(nm) can be conveniently obtained by
multiplying the ordinates by the electric field strength
$E$~(V~cm$^{-1}$) and dividing by Young's modulus $\ymod$~(kPa).

Figure~\ref{fig:1} shows $Z / E$ as a function of the scaled
$\zeta$-potential $\zeta e / (\kb T)$ for a particle with radius $a =
500$~nm and a hydrogel with Young's modulus $\ymod = 1$~kPa ($\prat =
1/2$) and Brinkman screening length $\bsl = 5$~nm. The electrolyte is
NaCl, with ionic strengths corresponding to $\ka = 1$--$10^3$. As
expected, the (negative) particle displacement is in the direction of
the electrical force, $4 \pi a^2 \sigma \vect{E}$, and increases with
the magnitude of the $\zeta$-potential ($\zeta < 0$).

At low $\zeta$-potentials, the displacement is clearly proportional to
$|\zeta|$ and, hence, the surface charge density $\sigma = \epsilon_s
\epsilon_o \kappa \zeta$ (when $|\zeta| < \kb T / e$). As expected,
the numerical solutions approach the boundary-layer theory
[Eqn.~(\ref{eqn:bltheory})] as $\ka \rightarrow \infty$. However, the
numerical results reveal distinct maximums at moderate and large
values of $|\zeta|$; these are due to polarization of the diffuse
double layer. As is well known from
electrophoresis~\citep[\eg,][]{OBrien:1978}, polarization diminishes
the local electric field, thereby attenuating the electrical
force. Because polarization by electro-migration and relaxation by
molecular diffusion are practically independent of the polymer gel in
this model~\citep{Hill:2006b}, they are as significant here as they
are in electrophoresis.

Figure~\ref{fig:2} shows $Z / E$ under the same conditions as in
figure~\ref{fig:1}, but now as a function of the scaled reciprocal
double-layer thickness $\ka$, with each curve corresponding to a
constant $\zeta$-potential. The displacement is clearly a weak
function of $\ka$, particularly when $|\zeta|$ is small, and, as
expected from figure~\ref{fig:1}, a much stronger function of
$\zeta$. Clearly, this way of presenting the results emphasizes the
large values of $\ka$ required for the boundary-layer theory to be
accurate.

Since figures~\ref{fig:1} and~\ref{fig:2} are presented with a fixed
value of $\bsl / a = 0.01$, it remains to establish the influence of
the Darcy permeability (or Brinkman screening length). Recall, the
boundary-layer theory [Eqn.~(\ref{eqn:bltheory})] indicates that the
displacement is independent of $\bsl$. More generally, however, the
particle displacement reflects the far-field electric-field-induced
distortion of the polymer skeleton when the particle is fixed at the
origin by an external force
\begin{equation}
  \vect{f}^E = -\vect{f}^{d,E} -\vect{f}^{e,E} -\vect{f}^{m,E} =
  -(8/3) \pi Z^E \ymod \vect{E}.
\end{equation}

\begin{figure}
  \begin{center}
    \vspace{1.0cm}
    \includegraphics[height=6cm]{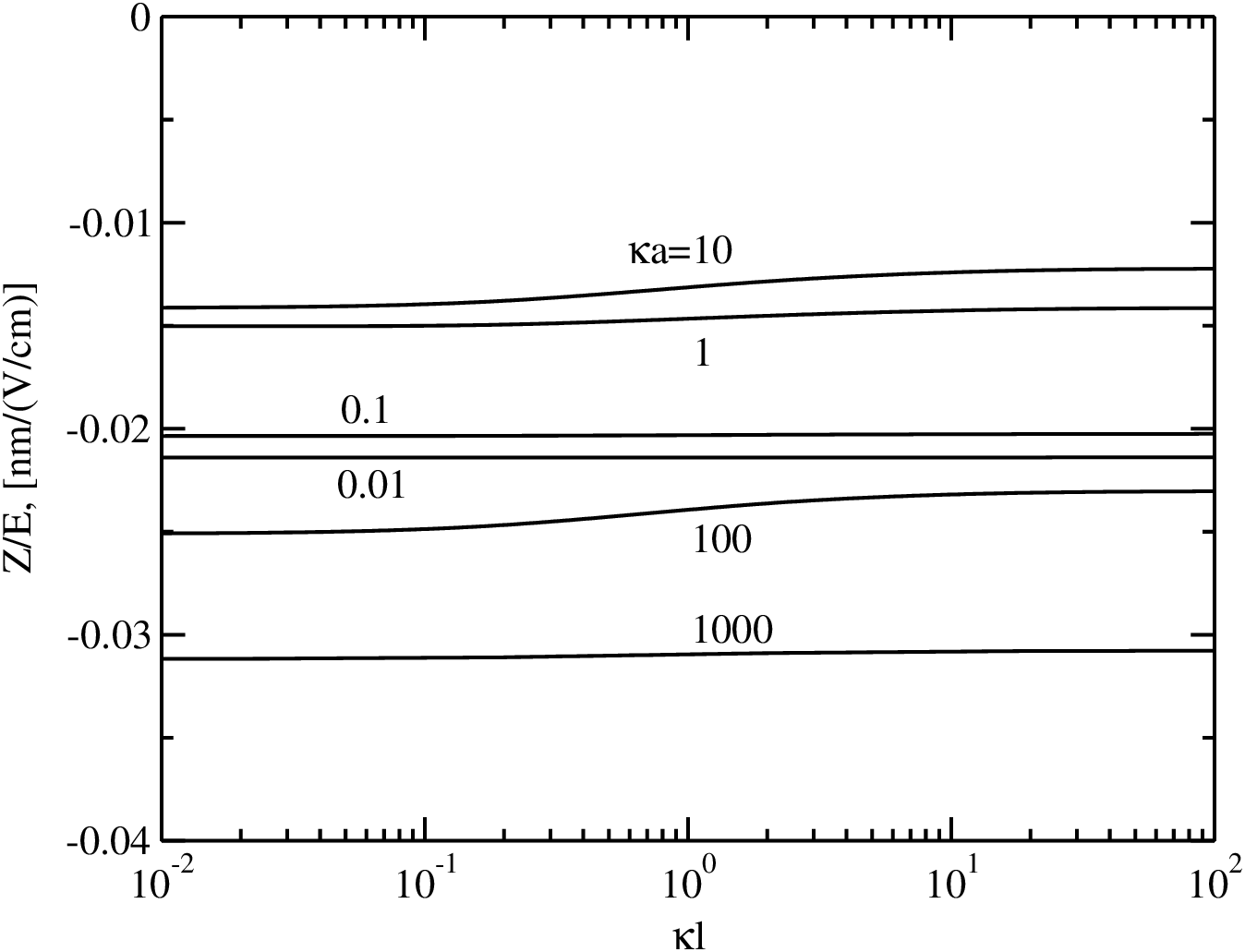}\\
    \vspace{0.5cm}
    \caption{\label{fig:2b} The ratio of the displacement $Z$ (nm) to
      the electric field strength $E$ (V~cm$^{-1}$) as a function of
      the scaled Brinkman screening length $\kappa \bsl$ with $\zeta e
      / (\kb T) = -6$ and $\ka = 0.01$, 0.1, 1,..., 1000: KCl at $T =
      298$~K; $a = 500$~nm; $\ymod = 1$~kPa; $\prat = 1/2$. Note that
      the displacement is inversely proportional to Young's modulus
      $\ymod$.}
  \end{center}
\end{figure}

Figure~\ref{fig:2b} shows how the particle displacement varies over
four decades of the scaled Brinkman screening length $\kappa \bsl$
when $\zeta e / (\kb T) = - 6$ and $\ka = 10^{-2}$--$10^3$. While
there are obvious transitions when $\kappa \bsl \sim 1$ (from plateaus
where $\kappa \bsl \rightarrow 0$ and $\infty$), the displacement is
remarkably insensitive to $\bsl$.

In the next section, we identify a simple transformation that permits
$Z^E$ and, hence, the particle displacement, to be approximated by the
well-known electrophoretic mobility, which, of course, is independent
of $\bsl$. Accordingly, we write
\begin{equation}
  \vect{Z} = 3 \zeta \epsilon_s \epsilon_o \ymod^{-1} f[\ka, \zeta e /
    (\kb T), \kappa \bsl] \vect{E},
\end{equation}
where the dimensionless function
\begin{eqnarray} \label{eqn:scaledz}
  f[\ka, \zeta e / (\kb T), \kappa \bsl] = \frac{4 \hat{Z}^E [\ka,
      \zeta e / (\kb T), \kappa \bsl]}{9 (\ka)^2 \zeta e / (\kb T)}
\end{eqnarray}
and, to a reasonable approximation,
\begin{equation}
f[\ka, \zeta e / (\kb T), \kappa \bsl] \approx f[\ka, \zeta e / (\kb
  T), \infty].
\end{equation}
It follows that figure~\ref{fig:1} or~\ref{fig:2} is sufficient to
  span a significant range of the parameter space (strictly for
  negatively charged inclusions and NaCl electrolyte). Furthermore,
  from Eqns.~(\ref{eqn:huckel}) and (\ref{eqn:bltheory}), it is
  evident that $f \rightarrow 2/3$ as $\ka \rightarrow 0$ (with
  $|\zeta| < \kb T / e$) and $f \rightarrow 1$ as $\ka \rightarrow
  \infty$.

\subsection{Connection to the electrophoretic mobility}

\begin{figure}
  \begin{center}
    \vspace{1cm}
    \subfigure[]{\includegraphics[width=5.5cm]{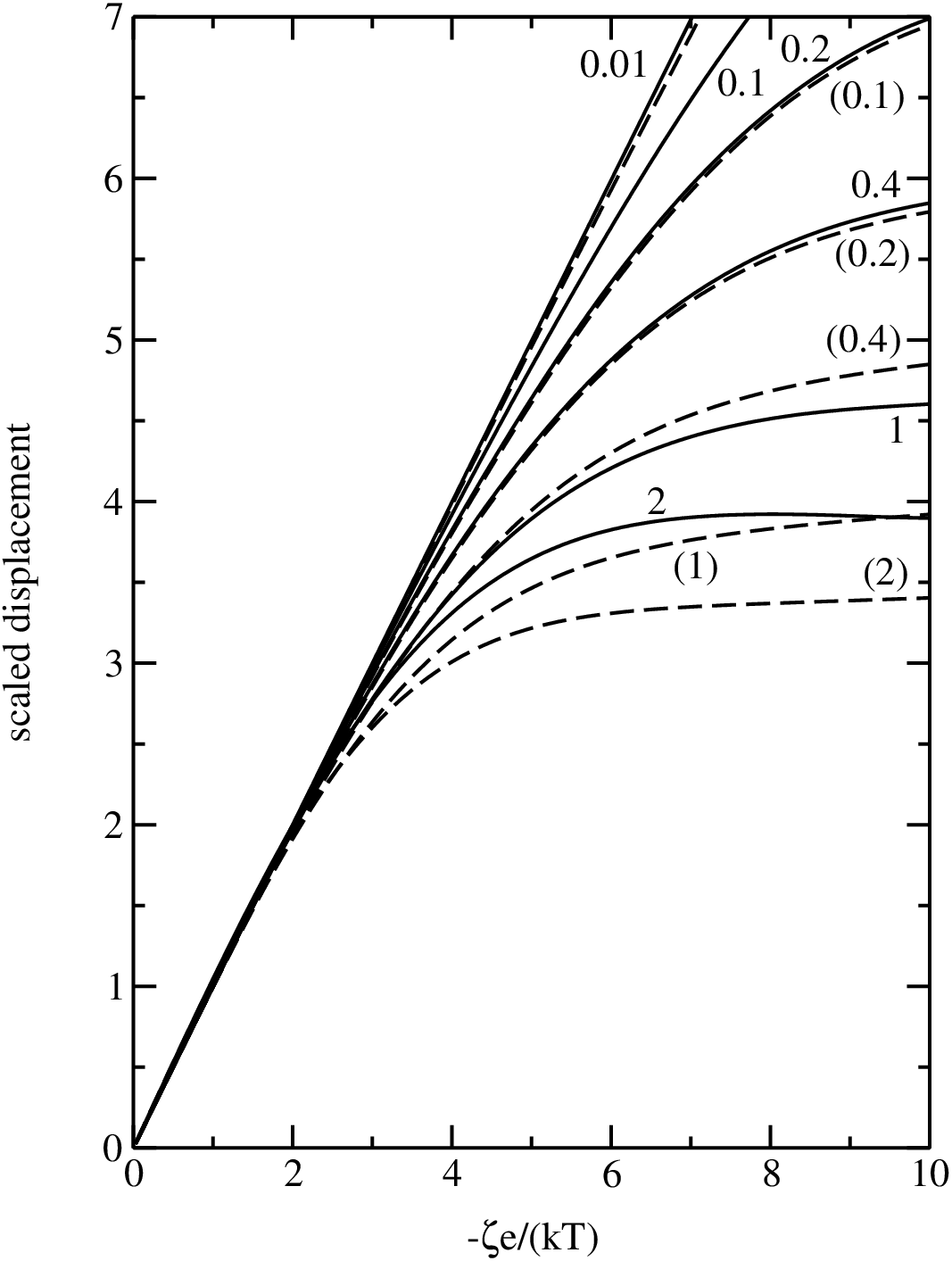}}
    \hspace{1cm}
    \subfigure[]{\includegraphics[width=5.5cm]{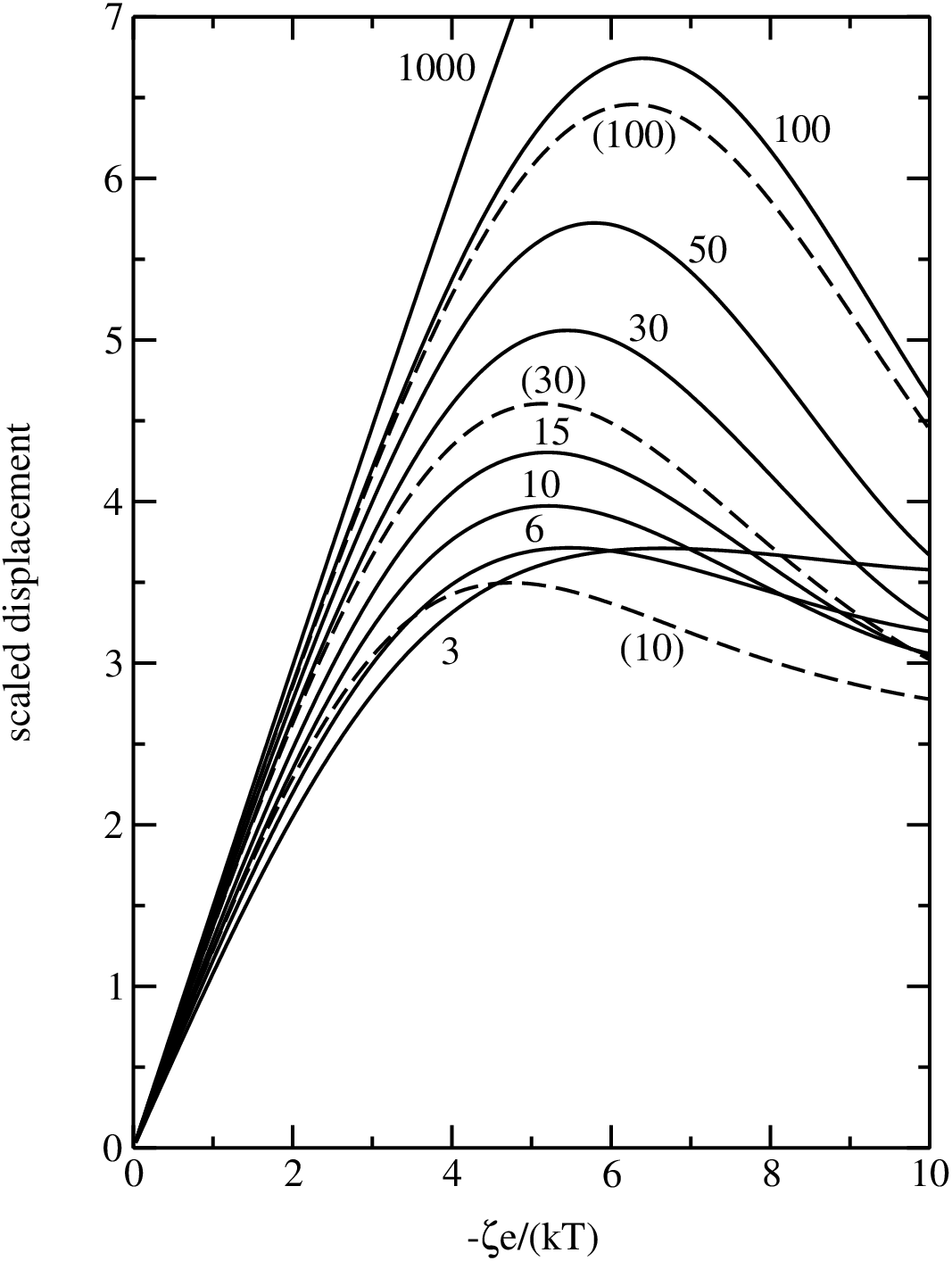}}
    \caption{\label{fig:3} The scaled displacement $-(Z/E) \ymod e
      / (2 \epsilon_s \epsilon_o \kb T)$ as a function of the scaled
      $\zeta$-potential $-\zeta e / (\kb T)$ for various scaled
      reciprocal double-layer thicknesses $\ka = 0.01$, 0.1, 0.2, 0.4,
      1 and 2 (solid lines, left panel); and $\ka = 3$, 6, 10, 15, 30,
      50, 100 and 1000 (solid lines, right panel): KCl at $T = 298$~K;
      $a = 500$~nm; $\bsl = 5$~nm; $\prat = 1/2$. Dashed lines are
      the scaled electrophoretic mobility $(U / E) 3 \eta e / (2
      \epsilon_s \epsilon_s \kb T)$ for selected values (labels in
      parentheses) of $\ka = 0.01$, 0.1, 0.2, 0.4, 1 and 2 (left
      panel); and 10, 30 and 100 (right panel).}
  \end{center}
\end{figure}

\begin{figure}
  \begin{center}
    \vspace{1.0cm} \includegraphics[width=6cm]{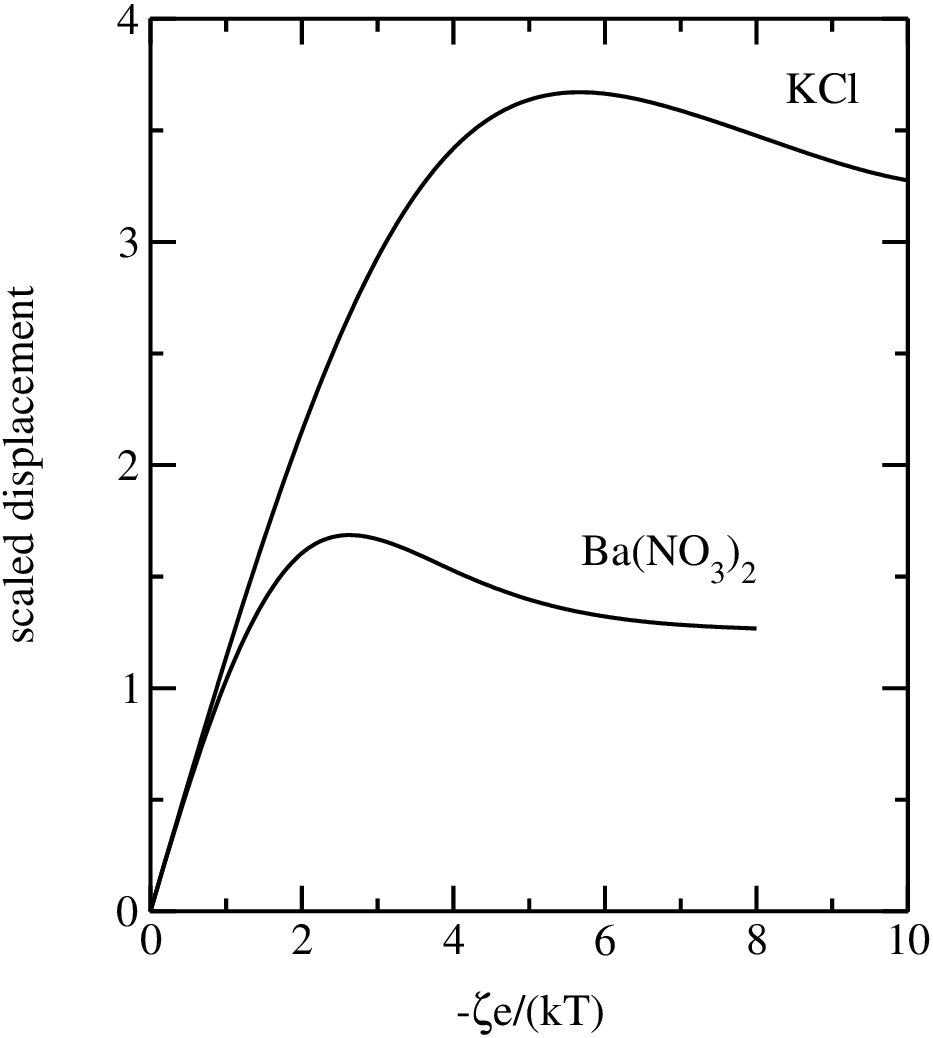}
    \vspace{0.5cm}
    \caption{\label{fig:4} The scaled displacement $-(Z/E) \ymod e
      / (2 \epsilon_s \epsilon_o \kb T)$ as a function of the scaled
      $\zeta$-potential $-\zeta e / (\kb T)$ for scaled reciprocal
      double-layer thickness $\ka = 5$: KCl and Ba(NO$_3$)$_2$ at $T =
      298$~K; $a = 500$~nm; $\bsl = 5$~nm; $\prat = 1/2$.}
  \end{center}
\end{figure}

The displacements shown in figures~\ref{fig:1} and~\ref{fig:2} bear a
close resemblance to the electrophoretic mobility~\citep{OBrien:1978},
so we present in figures~\ref{fig:3} and~\ref{fig:4} a {\em scaled
displacement}
\begin{equation}
  (Z/E) \ymod e/ (2 \epsilon_s \epsilon_o \kb T) = (3/2) f[\ka, \zeta
    e / (\kb T),\kappa \bsl] \zeta e/(\kb T)
\end{equation}
for a symmetrical electrolyte (KCl) (solid lines, figure~\ref{fig:3})
and a representative asymmetric electrolyte (Ba(NO$_3$)$_2$)
(figure~\ref{fig:4}). This dimensionless quantity (solid lines) has a
very similar dependence on $\ka$ and $\zeta e / (\kb T)$ as the {\em
scaled electrophoretic mobility} (dashed lines)
\begin{equation}
  (U / E) 3 \eta e / (2 \epsilon_s \epsilon_o \kb T)
\end{equation}
presented in O'Brien and White's well-known
paper~\citep[see][figures~3, 4, \& 6]{OBrien:1978}\footnote{The
electrophoretic mobilities reported here were calculated using
software (called MPEK, available from the corresponding author) based
on the methodology of Hill, Saville and Russel~\citep{Hill:2003a} for
the electrophoretic mobility and other single-particle properties of
spherical polymer-coated colloids. Here, the influence of a polymer
coating is removed by specifying an infinite (large) Darcy
permeability for the coating. The mobilities in figures~\ref{fig:3}
and~\ref{fig:4} are in excellent agreement with O'Brien and White's
calculations~\citep{OBrien:1978}; small differences can be attributed
to the opposite sign of the $\zeta$-potential.}. Here, $U$ is the
electrophoretic velocity, \ie, the translational velocity acquired by
a charged spherical colloid dispersed in a Newtonian electrolyte in
response to a (weak) electric field $E$.

The difference between the scaled mobility and scaled displacement
highlighted in figure~\ref{fig:3} is small when $|\zeta| < \kb T / e$,
but increases appreciably when $\ka \sim 1$ and $|\zeta| > 3 \kb T /
e$. Note that the scaled mobility is smaller than the respective
scaled displacement, because the electrophoresis problem
over-estimates the convective contribution to the ion fluxes, thereby
over-polarizing the diffuse double layer and, therefore,
under-estimating the electrical force in the particle-displacement
problem.

A simple but approximate relationship between $Z / E$ and $U / E$ can
be established by eliminating the Darcy drag force from the fluid
momentum equation [Eqn.~(\ref{eqn:nseqn})] and the polymer equation of
static equilibrium [Eqn.~(\ref{eqn:staticeqm})]. This produces the
momentum conservation equation in the electrophoresis problem [\ie,
Eqn.~(\ref{eqn:nseqn}) without the Darcy drag term] with a modified
`fluid' velocity
\begin{equation} \label{eqn:modvel}
  \vect{u}'= \vect{u} + \vect{v} \ymod/(3 \eta).
\end{equation}
Recall, $\vect{u}$ is the fluid velocity in the polymer gel and
$\vect{v}$ is the displacement of the skeleton.

By setting $\vect{u} = \vect{u}'$ in the ion conservation equation,
the solution of the electrophoresis problem (involving only
$\vect{u}'$) over-estimates the convective ion fluxes by an
$O[(v_c/u_c) \ymod / (3 \eta)]$ amount; here, $v_c$ and $u_c$ are,
respectively, characteristic polymer displacement and fluid velocity
scales. However, since the convective ion fluxes are $O(\Pe_j)$, where
the P{\'e}clet numbers $\Pe_j = u_c \kappa^{-1} / D_j \ll 1$, the
absolute errors are small.

Note that the polymer displacement reflects a transfer of the
electrical body force from the fluid to the elastic skeleton. This
transfer occurs by direct coupling of the fluid and polymer (via the
Darcy drag force) and through indirect coupling by the transfer of
viscous stresses from the fluid to the particle, which, in turn, are
transferred to the polymer via mechanical contact between the particle
and polymer. Consequently, the polymer distortion must be independent
of the permeability in so far as the electrical body force is
constant. However, the electrically driven flow increases
significantly with the permeability [either as $\bsl$ or $\bsl^2$,
depending on $\kappa^{-1}$~\citep{Hill:2006b,Hill:2006c}], so the
$O[(v_c/u_c) \ymod / (3 \eta)]$ errors in the ion conservation
equations must diminish with increasing permeability. Accordingly, the
solution of the electrophoresis problem must yield an increasingly
accurate solution of the particle displacement problem as $\kappa \bsl
\rightarrow \infty$.

\begin{figure}
  \begin{center}
    \vspace{1cm}
    \subfigure[]{\includegraphics[width=5.5cm]{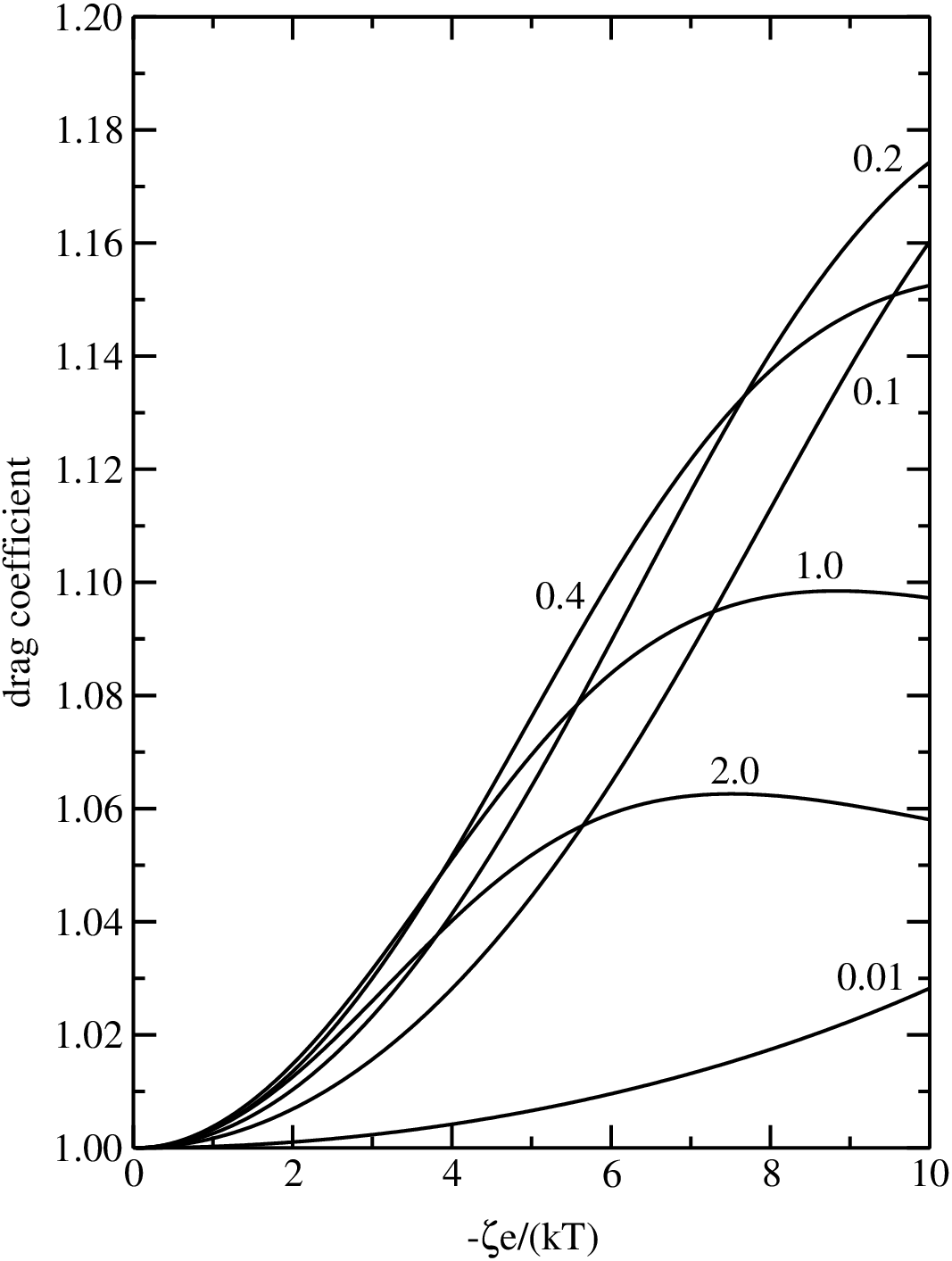}}
    \hspace{1cm}
    \subfigure[]{\includegraphics[width=5.5cm]{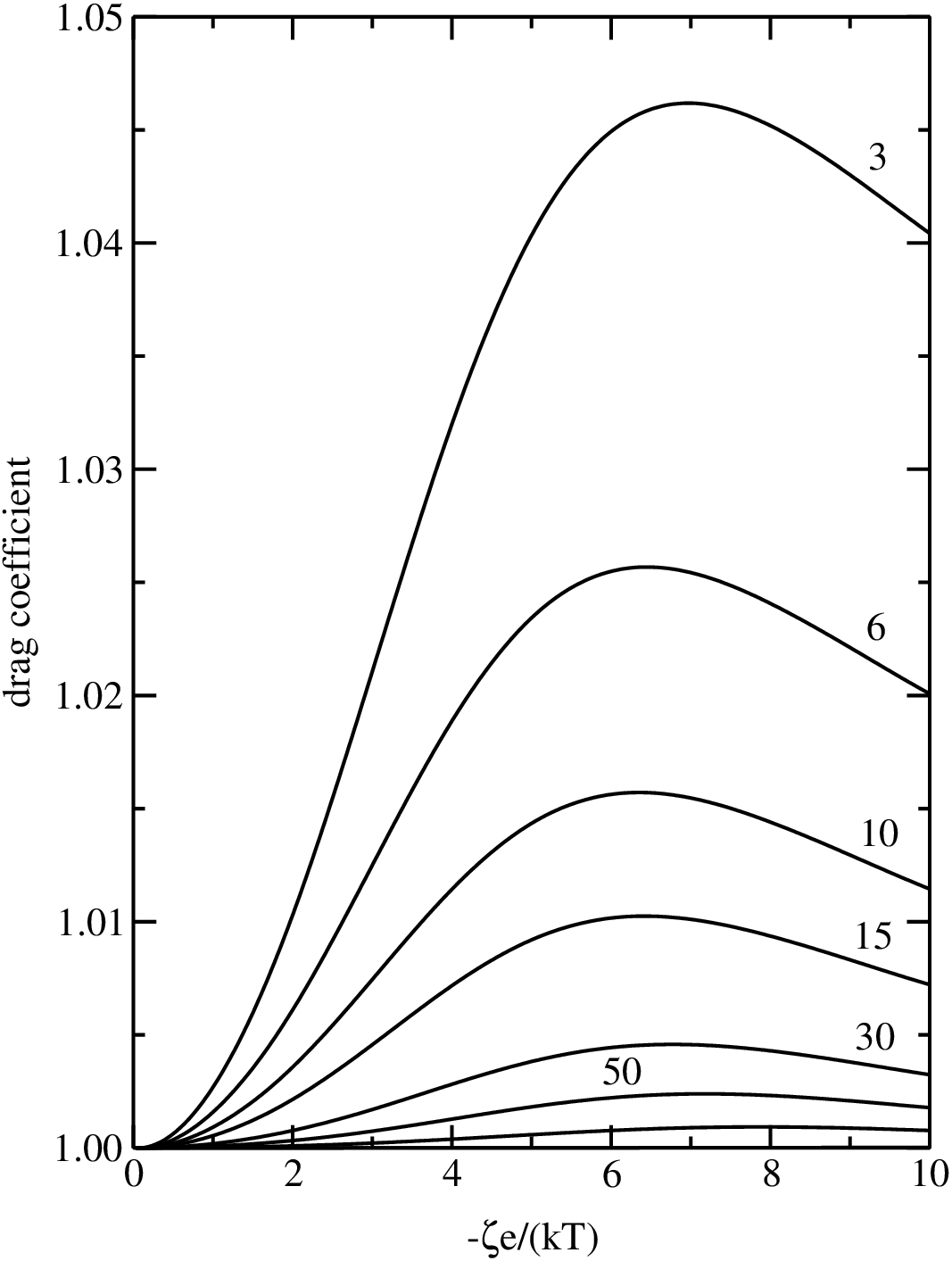}}
    \caption{\label{fig:drag} The drag coefficient $F = (4/3)
      \overline{C}^U / a = D_s / D[\ka, \zeta e / (\kb T)]$ for
      charged spherical colloids as a function of the scaled
      $\zeta$-potential $-\zeta e / (\kb T)$ for various scaled
      reciprocal double-layer thicknesses $\ka = 0.01$, 0.1, 0.2, 0.4,
      1 and 2 (left panel); and $\ka = 3$, 6, 10, 15, 30, 50, 100 and
      1000 (right panel): KCl at $T = 298$~K.}
  \end{center}
\end{figure}

More quantitatively, the solution of the (E) electrophoresis
problem~\citep{OBrien:1978} yields $u' \sim \overline{C}^E E r^{-1}$
as $r \rightarrow \infty$. Therefore, because $v \sim Z^E E r^{-1}$
and $u \sim C^E E r^{-3}$ as $r \rightarrow \infty$, it follows from
Eqn.~(\ref{eqn:modvel}) that
\begin{equation} \label{eqn:zapprox}
  Z^E \rightarrow \overline{C}^E 3 \eta / \ymod \mbox{ as } \kappa
  \bsl \rightarrow \infty \ \ (\prat = 1/2).
\end{equation}
Furthermore, since the electrophoretic mobility
\begin{equation} \label{eqn:mobility}
  M \equiv U / E = \overline{C}^E / \overline{C}^U,
\end{equation}
where $\overline{C}^E$ and $\overline{C}^U$ are the asymptotic
coefficients associated with the far-field decay of the fluid velocity
in the (E) and (U) (electrophoresis) problems~\citep{OBrien:1978},
Eqns.~(\ref{eqn:displacement}),~(\ref{eqn:zapprox})
and~(\ref{eqn:mobility}) give
\begin{eqnarray} \label{eqn:zapprox2}
  \vect{Z} \rightarrow F (3 \eta /
    \ymod) M \vect{E} \mbox{ as } \kappa \bsl \rightarrow \infty \ \
    (\prat = 1/2).
\end{eqnarray}
Note that $F = (4/3) \overline{C}^U / a = D_s / D \sim 1$ is the
particle {\em drag coefficient}, which can be conveniently expressed
as the ratio of the {\em Stokes-Einstein-Sutherland}
diffusivity\footnote{\cite{Squires:2005} present a compelling case to
associate W.~Sutherland with this famous relationship.} $D_s = \kb T /
(6 \pi \eta a)$ to the actual diffusivity $D[\ka, \zeta e / (\kb T)]
\ge D_s$. The drag coefficient is shown in figure~\ref{fig:drag} for
negatively charged colloidal spheres in a KCl electrolyte. Since $1
\le F < 1.2$ and most often $1 \le F \lesssim 1.02$, the error in the
particle displacement that comes from setting $F = 1$ in
Eqn.~(\ref{eqn:zapprox2}) tends to be small compared to the error
arising from finite $\kappa \bsl$.

To demonstrate the correctness of Eqn.~(\ref{eqn:zapprox2}), let us
briefly consider a specific example where $\zeta e / (\kb T) = -6$ and
$\ka = 10$. From figure~\ref{fig:2b}, the scaled displacements are
$\approx 3.89$ and $\approx 3.42$ when $\kappa \bsl = 0.1$ and $\kappa
\bsl \rightarrow \infty$, respectively. Furthermore, from
figure~\ref{fig:3}, the scaled mobility and drag coefficient are
$\approx 3.36$ and $\approx 1.016$, respectively. Therefore,
multiplying the scaled mobility by the drag coefficient gives $\approx
3.41$, which, as expected, compares extremely well with the value
($\approx 3.42$) obtained directly when $\kappa \bsl \gg 1$.

In summary, we have established that the electrophoretic mobility and
drag coefficient combine [as indicated in Eqn.~(\ref{eqn:zapprox2})]
to yield the correct limiting value of the scaled particle
displacement as $\kappa \bsl \rightarrow \infty$. Furthermore, the
difference between the displacement with impenetrable and infinitely
permeable polymer reflects the degree to which polymer influences
convective ion transport. Finally, because polarization by convection
is weak relative to electromigrative polarization and diffusive
relaxation, the influence of permeability on the displacement is
generally small. It should be noted, however, that Young's modulus (or
shear modulus) of the skeleton and the permeability both vary with the
polymer density according to scaling laws that have been studied
extensively in the polymer physics
literature~\citep[\eg,][]{deGennes:1979}. In practice, therefore,
particle displacements are expected to vary primarily with the
modulus, with relatively insignificant changes due to the accompanying
change in permeability.

\section{Application to micro-rheology}

As demonstrated above, dimensional particle displacements are
$\lesssim 10 \epsilon_s \epsilon_o \kb T E / (e \ymod)$ in the
experimentally accessible range of $\ka$ and $|\zeta| e / (\kb T)$. In
polymer gels saturated with aqueous electrolytes at room temperature,
this implies that $Z < 0.17 (E / \ymod)$~nm. For example, with $\ymod
= 10$~kPa and $E = 100$~V~cm$^{-1}$, we find $Z < 0.17$~nm, which
poses a significant, but not necessarily insurmountable, experimental
challenge. When $\ymod = 0.1$~kPa, however, $Z < 1.7$~nm is achieved
with $E = 10$~V~cm$^{-1}$.

As highlighted in the introduction, passive micro-rheology also tends
to be limited to complex fluids whose storage modulus is $\lesssim
1$~kPa~\citep[\eg,][for F-actin solutions and polyacrylamide gels, and
$\lambda$-DNA, respectively]{Schnurr:1997,Chen:2003}. For hydrogels,
such conditions are realized when the cross-linking density is
low~\citep{deGennes:1979}. As our theory demonstrates, high electric
field strengths ($\gtrsim 1$~kV~cm$^{-1}$) and/or low elastic moduli
($\lesssim 1$~kPa) are necessary to register nanometer-scale
electric-field-induced particle displacements. Accordingly, to
minimize the electric field strength and, hence, avoid non-linear
electrokinetic phenomena and Joule heating, we propose an experimental
methodology that seeks to resolve sub-nanometer particle
displacements.

In contrast to existing laser tweezers micro-rheology, where two
lasers are used to simultaneously perturb and image a micron-sized
tracer particle~\citep{Valentine:1996,Yamaguchi:2005}, we envision an
experiment that necessitates only an imaging laser to measure $Z / E =
(4/3) (Z^E/a)$, which, recall, reflects physicochemical
characteristics of the particle-gel interface. In practice, an optical
trap and piezo-electric stage (with feed-back control) may be adopted
for positioning and alignment purposes.

Very small-amplitude particle displacements induced by an oscillatory
electric field may be registered by correlating the measured response
$Z' = Z - \langle Z \rangle = Z''(t) + |\hat{Z}| \cos{(\omega t +
\theta)}$ with the (harmonic) electric field strength $E = |\hat{E}|
\cos{(\omega t)}$. Here, $Z''(t)$ denotes the noise (adjusted to have
zero mean), $|\hat{Z}|$ and $|\hat{E}|$ are the amplitudes of the
(harmonic) particle displacement and electric field, $\theta$ is the
phase angle, $\omega$ is the angular frequency, and $t$ is time. If
the noise $Z''$ is uncorrelated with the excitation $E$, then
\begin{equation}
  \langle Z' E \rangle \equiv T^{-1} \int_0^T Z'(t) E(t) \mbox{d}t
  \rightarrow (1/2) |\hat{Z}||\hat{E}| \cos{\theta} + O(T^{-1/2})
\end{equation}
if the duration $T$ of the time series is large.

The quasi-steady particle equation of motion [Eqn.~(\ref{eqn:fbal})
with viscous drag force $-6 \pi \eta^* a \mbox{d} \vect{Z} /
\mbox{d}t$~\citep{Valentine:1996,Yamaguchi:2005}] yields a (complex)
displacement
\begin{equation}
  \hat{Z} = \frac{\hat{E} (4/3) (Z^E/a)}{(1 + i \omega 3 \eta^* /
    \ymod)} \ \ (\prat = 1/2),
\end{equation}
with amplitude and phase given by
\begin{equation}
  |\hat{Z}| = \frac{|\hat{E}| (4/3) (Z^E/a)}{\sqrt{1 + (\omega 3 \eta^*
      / \ymod)^2}} \ \mbox{ and } \ \tan{\theta} = - \omega 3 \eta^*
      / \ymod.
\end{equation}
Therefore, if the viscosity and elastic modulus are known, from a
complementary macro- or micro-rheology experiment, then
\begin{equation}
  (4/3) (Z^E/a) = \frac{\langle Z' E \rangle}{(1/2) \cos{\theta}
  |\hat{E}|^2 \sqrt{1 + (\tan{\theta})^2}}.
\end{equation}
Moreover, at sufficiently low frequencies (\ie, $\omega \ll \ymod / (3
\eta^*)$), the quasi-steady response is simply
\begin{equation}
  (4/3) (Z^E/a) \approx 2 \langle Z' E \rangle / |\hat{E}|^2 \ (\prat
  = 1/2).
\end{equation}
In this manner, it is not necessary to have explicit knowledge of the
visco-elastic characteristics of the gel to directly probe the
physicochemical characteristics of the particle, as characterized by
$(4/3) (Z^E/a)$. Furthermore, as a compelling alternative to a more
direct measurement of $Z/E$ to determine $(4/3) (Z^E/a)$, the
correlation $\langle Z' E \rangle$ may provide an effective means of
dealing with the small signal-to-noise ratio expected in such an
experiment.

Finally, when the frequency is higher than the reciprocal draining
time [Eqn.~(\ref{eqn:draining})], a non-trivial extension of the
quasi-steady theory is necessary to account for a compressible polymer
skeleton and oscillatory relative motion of the fluid and polymer
phases. \cite{Levine:2001} undertook such a study for particles
embedded in polymer gels without electrokinetic influences. They
significantly extended the prevailing quasi-steady
theory~\citep[\eg,][]{Levine:2001b}, identifying the frequencies
where, in particular, compressibility of the polymer skeleton
influences the dynamical response. Clearly, their work provides a
sound basis for advancing the present theory to elucidate dynamical
electrokinetic effects in micro-rheology applications.

\section{Summary}

We presented a theoretical model to calculate the
electric-field-induced displacement of a charged, spherical colloid
embedded in an electrolyte-saturated polymer gel. The standard
electrokinetic model describes electric-field-induced transport of
ions and electrolyte momentum, with a Darcy-drag term that couples the
electrolyte mass and momentum conservation equations (Brinkman's
equations) to a continuum equation of static equilibrium for an
unbounded, incompressible, linearly elastic polymer skeleton.

The scaled particle displacement \[(Z/E) \ymod e / (2 \epsilon_s
\epsilon_o \kb T)\] has a similar dependence on $\zeta e / (\kb T)$
and $\ka$ as the well-known scaled electrophoretic mobility
\begin{equation}
  (U/E) 3 \eta e (2 \epsilon_s \epsilon_o \kb T). \nonumber
\end{equation}
More precisely, we showed that the product of the scaled
electrophoretic mobility and particle drag (friction) coefficient
yields the exact scaled displacement when $\kappa \bsl \rightarrow
\infty$ (\ie, when the polymer skeleton presents zero hydrodynamic
resistance to flow). However, because the particle displacement
decreases only slightly as $\kappa \bsl$ passes through $\sim 1$, when
increasing from 0 to $\infty$, the scaled electrophoretic mobility
provides an excellent approximation of the scaled displacement over a
wide range of the experimentally accessible parameter
space. Therefore, as expected from electrophoresis, the scaled
displacement is linear in the $\zeta$-potential when $|\zeta| \lesssim
2 \kb T / e$, and has distinct maximums when $|\zeta| \sim 5 \kb T /
e$. The decrease in displacement with increasing $\zeta$-potential is
due to polarization and relaxation of the diffuse double
layer. Evidently, polarization is dominated by electromigration, since
the influence of convection, which can be completely arrested by an
hydrodynamically impenetrable polymer skeleton, is extremely weak.

To help deal with the small signal-to-noise ratio expected in
experiments that necessitate sub-nanometer particle displacements, we
established a connection between $Z / E = (4/3) (Z^E/a)$, which
reflects the physicochemical state of the particle-gel interface, and
the correlation of the response ($Z'$) and forcing ($E$) signals in a
back-focal-plane interferometry apparatus. As a particle
characterization tool, the novel micro-electrokinetic experiment we
describe is analogous to classical micro-electrophoresis, with
existing laser tweezer micro-rheology serving as the (complementary)
analogue of dynamic light scattering.

We did not examine the stresses in the surrounding
polymer. Nevertheless, our calculations provide an important first
step toward future studies aimed at quantifying the micro-scale states
of stress (and strain) when these soft composite materials are
subjected to electric fields. Our model may be helpful for studying
fracture, and it provides a means of quantitatively interpreting
experiments designed to measure small electric-field-induced
displacements of charged inclusions. In turn, these could be used to
probe the mechanics of weak (uncharged) polymer gels at length and
time scales that are beyond the reach of conventional (macroscale)
rheometers.


\begin{acknowledgements}
  RJH gratefully acknowledges support from the Natural Sciences and
  Engineering Research Council of Canada (NSERC) (grant number 204542)
  and the Canada Research Chairs program (Tier II).
\end{acknowledgements}

\bibliography{/home/rhill/latex/bibliographies/global}

\appendix

\section{The force to displace a finite sized sphere embedded in a compressible
  elastic continuum} \label{app:1}

This appendix provides an analytical solution of the equation of
static equilibrium in the absence of body forces (Darcy drag). In
turn, the force required to displace (by distance $\vect{Z}$) a rigid
sphere (with radius $a$) embedded in an unbounded elastic continuum
(with Young's modulus $\ymod$ and Poisson's ratio $\prat$) is
obtained.

Substituting a solution of the form
\begin{equation}
  \vect{v} = \vect{v}_0 + \vect{v}_1,
\end{equation}
where
\begin{equation}
  \lapl{\vect{v}_0} = \vect{0},
\end{equation}
into the equation of static equilibrium [Eqn.~(\ref{eqn:staticeqm})]
gives
\begin{eqnarray} \label{eqn:mix}
  \lapl{\vect{v}_1} + \frac{1}{(1 - 2 \prat)} \grad{[\dive{(\vect{v}_0
  +\vect{v}_1)}]} = \vect{0}.
\end{eqnarray}
Taking the curl yields
\begin{equation} \label{eqn:vort}
  \lapl{(\curl{\vect{v}_1})} = 0,
\end{equation}
which, with the prevailing axial symmetry, provides a scalar equation
for $(\curl{\vect{v}_1}) \cdot \vect{e}_\phi$. Since
$\curl{\vect{v}_1}$ is an harmonic pseudo vector, symmetry and
linearity yield a non-zero decaying solution $\curl{\vect{v}_1} \sim
\vect{Z} \times \grad{r^{-1}}$. It follows that $\vect{v}_1 \sim
\vect{Z} r^{-1}$, which is harmonic and, hence, can be attributed to
$\vect{v}_0$. Accordingly, the only non-harmonic contribution to
$\vect{v}_1$ is irrotational ($\curl{\vect{v}_1} = \vect{0}$) and,
hence,
\begin{equation}
  \vect{v}_1 = \grad{\phi},
\end{equation}
where $\phi$ is a scalar function of position. Substituting this into
Eqn.~(\ref{eqn:mix}) gives
\begin{equation}
  \grad{[\dive{\vect{v}_0} + 2 (1 - \prat) \lapl{\phi}]} =
  \vect{0},
\end{equation}
so
\begin{equation}
  \dive{\vect{v}_0} + 2 (1 - \prat) \lapl{\phi} = 0.
\end{equation}
Again, symmetry and linearity considerations yield the general
decaying solution
\begin{equation}
  \vect{v}_0 = c_0 \vect{Z} + c_1 \vect{Z} r^{-1} + c_2 (\vect{Z}
  \cdot \grad) \grad (r^{-1}).
\end{equation}
It follows that
\begin{equation}
  \dive{\vect{v}_0} = c_1 \vect{Z} \cdot \grad(r^{-1}) \ \ (= - c_1
  \vect{Z} \cdot \vect{e}_r r^{-2})
\end{equation}
and, hence,
\begin{equation} \label{eqn:mix1}
  \lapl{\phi} = - \frac{c_1 \vect{Z} \cdot \grad(r^{-1})}{2 (1 - \prat)}.
\end{equation}
Note that $\lapl{\psi} = r^{-1}$ has the solution $\psi = r / 2$, so
writing Eqn.~(\ref{eqn:mix1}) as
\begin{equation}
  \lapl{\phi} = - \frac{c_1 \vect{Z} \cdot \grad{(\lapl{\psi})}}{2 (1
    - \prat)}
\end{equation}
requires
\begin{equation}
  \phi = -\frac{c_1 \vect{Z} \cdot \vect{e}_r}{4 (1 - \prat)}
\end{equation}
and, hence,
\begin{equation}
  \vect{v}_1 = \grad{\phi} = -\frac{c_1}{4 r (1 - \prat)} [\vect{Z} -
    (\vect{Z} \cdot \vect{e}_r) \vect{e}_r].
\end{equation}
Finally, the complete displacement field is
\begin{eqnarray} \label{eqn:full}
  \vect{v} = c_0 \vect{Z} + \frac{c_1}{r} \vect{Z} - \frac{c_1}{4 r (1
    - \prat)} [\vect{Z} - (\vect{Z} \cdot \vect{e}_r) \vect{e}_r] +
  \frac{c_2}{r^3} [3 (\vect{Z} \cdot \vect{e}_r) \vect{e}_r -
    \vect{Z}],
\end{eqnarray}
where the scalar constants $c_0$, $c_1$ and $c_2$ must be chosen to
satisfy the boundary conditions.

\subsection{No-slip boundary condition}

If, for example, $\vect{v} \rightarrow - \vect{Z}$ as $r \rightarrow
\infty$ and $\vect{v} = \vect{0}$ at $r = a$ (fixed), then $c_0 = -1$
and
\begin{eqnarray}
  \vect{0} = - \vect{Z} + \frac{c_1}{a} \vect{Z} + \frac{c_2}{a^3} [3
    (\vect{Z} \cdot \vect{e}_r) \vect{e}_r - \vect{Z}] - \frac{c_1}{4
    a (1 - \prat)} [\vect{Z} - (\vect{Z} \cdot \vect{e}_r)
    \vect{e}_r],
\end{eqnarray}
which requires
\begin{equation}
  c_1 = \frac{6 (1-\prat) a}{(5 - 6 \prat)} \mbox{ and } c_2 =
  \frac{a^3}{2 (6 \prat - 5)}.
\end{equation}

The mechanical-contact force on the inclusion is therefore
\begin{eqnarray} \label{eqn:elasticforce}
  \vect{f}^m = \int_{r=a} \vect{\sigma} \cdot \hat{\vect{n}} \mbox{d}A
  = - \frac{2 \pi \ymod c_1 \vect{Z}}{(1 + \prat)} = - \frac{12 \pi
    \ymod a \vect{Z} (1-\prat)}{(5 - 6 \prat)(1 + \prat)}.
\end{eqnarray}

Note that the displacement field can be rewritten in terms of the
force, so the Green's function
\begin{equation}
  \vect{G} = \frac{(1+\prat)}{8 \pi \ymod (1 - \prat) r} [(3
    - 4 \prat) \vect{\delta} + \vect{e}_r\vect{e}_r]
\end{equation}
is obtained by changing reference frames ($\vect{v} = \vect{Z}$ at $r
= a$, and $\vect{v} \rightarrow \vect{0}$ as $r \rightarrow \infty$)
and letting $a \rightarrow 0$.

\subsection{Slip boundary condition}

Again, if $\vect{v} \rightarrow - \vect{Z}$ as $r \rightarrow \infty$,
but $\vect{v} \cdot \vect{e}_r = 0$ at $r = a$ (zero radial
displacement) and $\vect{\sigma} \cdot \vect{e}_r - (\vect{\sigma}
\cdot \vect{e}_r) \cdot \vect{e}_r \vect{e}_r = \vect{0}$ at $r = a$
(zero tangential stress), then $c_0 = -1$, and
\begin{equation}
  c_1 = \frac{6 (1 - \prat) a}{(7 - 8 \prat)} \mbox{ and } c_2 =
  \frac{(1 - 2 \prat) a^3}{2 (7 - 8 \prat)}.
\end{equation} 
The mechanical-contact force (on the inclusion) is then
\begin{eqnarray} \label{eqn:elasticforceslip}
  \vect{f}^m = - \frac{2 \pi \ymod c_1 \vect{Z}}{(1 + \prat)} = -
  \frac{12 \pi \ymod a \vect{Z} (1 - \prat)}{(7 - 8 \prat)(1 +
  \prat)}.
\end{eqnarray}

\section{The leading-order isotropic stress for an incompressible
  elastic continuum} \label{app:elastic}

Writing the displacement field as
\begin{equation}
  \vect{v} = \vect{v}_0 + \epsilon \vect{v}_1 + \epsilon^2 \vect{v}_2
  + ...,
\end{equation}
where $\epsilon = (1-2\prat) \ll 1$, substituting this into the
equation of static equilibrium
\begin{equation} \label{eqn:static}
  \lapl{\vect{v}} + \frac{1}{\epsilon} \grad{(\dive{\vect{v}})} = -
  (\eta / \bsl^2) \vect{u} \frac{(3 - \epsilon)}{\ymod},
\end{equation}
and collecting terms of like order in $\epsilon$ gives at $O(1)$:
\begin{equation} \label{eqn:oone}
  \grad{(\dive{\vect{v}_0})} = \vect{0};
\end{equation}
at $O(\epsilon)$
\begin{equation} \label{eqn:oepsilon}
  \lapl \vect{v}_0 + \grad{(\dive{\vect{v}_1})} = - (\eta / \bsl^2)
  \vect{u} \frac{3}{\ymod};
\end{equation}
at $O(\epsilon^2)$
\begin{equation}
  \lapl{\vect{v}_1} + \grad{(\dive{\vect{v}_2})} = (\eta / \bsl^2)
  \vect{u} \frac{1}{\ymod};
\end{equation}
and at $O(\epsilon^3)$
\begin{equation}
  \lapl{\vect{v}_2} + \grad{(\dive{\vect{v}_3})} = \vect{0}.
\end{equation}
Note that, if $\vect{u}$ and $\vect{v}_0$ are both divergence-free,
then $\dive{\vect{v}_1}$, $\dive{\vect{v}_2}$, \etc, all satisfy
Laplace's equation with general solution, \eg, $\dive{\vect{v}_1} =
a_1 + b_1 \vect{E} \cdot \grad{r^{-1}}$.

Because the leading contribution to the displacement $\vect{v}_0$ is
divergence-free [Eqn.~(\ref{eqn:oone})], it can be written
\begin{equation}
  \vect{v}_0 = \curl \curl g(r) \vect{E}.
\end{equation}
Substituting this into the curl of
Eqn.~(\ref{eqn:oepsilon}) gives
\begin{equation}
  \frac{\mbox{d}}{\mbox{d}r} \lapl \lapl g + \frac{3 \eta}{\ymod
  \bsl^2} \frac{\mbox{d}}{\mbox{d}r} \lapl f = 0,
\end{equation}
where
\begin{equation}
  \lapl = \frac{1}{r^2} \frac{\mbox{d}}{\mbox{d}r} (r^2
  \frac{\mbox{d}}{\mbox{d}r})
\end{equation}
and $f(r)$ is known. Clearly, the solution is independent of
$\vect{v}_1$. Note, however, that $\vect{v}_1$ contributes to the
leading-order stress,
\begin{equation}
  \vect{\sigma}_0 = \frac{2 \ymod}{3} [\vect{e}_0 + \frac{1}{2}
    (\dive{\vect{v}_1}) \vect{\delta}],
\end{equation}
where $\vect{e}_0 = (1/2) [\grad{\vect{v}_0} +
  (\grad{\vect{v}_0})^T]$. It follows that the leading contribution to
the integral of the surface traction is
\begin{equation}
  \vect{f}'_0 = \frac{\ymod}{3} \int_{r\rightarrow \infty}
       [\grad{\vect{v}_0} + (\grad{\vect{v}_0})^T +
	 (\dive{\vect{v}_1}) \vect{\delta}] \cdot \vect{e}_r
       \mbox{d}A,
\end{equation}
where [Eqn.~(\ref{eqn:oepsilon})]
\begin{equation}
  \dive{\vect{v}_1} = -\int_{\infty}^r [(\eta / \bsl^2) \vect{u}
    \frac{3}{\ymod} + \lapl \vect{v}_0] \cdot \vect{e}_r
  \mbox{d}r'.
\end{equation}

Note that only the far-field decays of $\vect{u}$ and $\vect{v}_0$ are
necessary to evaluate this integral when $r \rightarrow
\infty$. Recall,
\begin{equation}
  \vect{u} \rightarrow - 2 C^E r^{-3} (\vect{E} \cdot \vect{e}_r)
  \vect{e}_r - C^E r^{-3} (\vect{E} \cdot \vect{e}_\theta)
  \vect{e}_\theta \mbox{ as } r \rightarrow \infty,
\end{equation}
so
\begin{equation}
  \dive{\vect{v}_1} \rightarrow [2 Z^E - (\eta / \bsl^2)
  \frac{3}{\ymod} C^E] r^{-2} (\vect{E} \cdot \vect{e}_r) \mbox{ as
  } r \rightarrow \infty
\end{equation}
and, hence,
\begin{eqnarray}
  \vect{f}'_0 &=& \frac{\ymod}{3} \int_{r \rightarrow \infty} \left
    [ \left (\frac{\partial v_{0,i}}{\partial x_j} + \frac{\partial
    v_{0,j}}{\partial x_i} \right) + \frac{\partial v_{1,k}}{\partial
    x_k} \delta_{ij} \right ] e_{r,j} \mbox{d}A \nonumber \\ &=& \ymod (16/9) \pi Z^E \vect{E} + \ymod (8/9) \pi Z^E \vect{E} -
    (4/3) \pi(\eta / \bsl^2) C^E \vect{E} \nonumber \\ &=& (8/3) \pi
    Z^E \ymod \vect{E} - (4/3) \pi (\eta / \bsl^2) C^E \vect{E}
    \end{eqnarray}
Finally, adding the volume integral [$-(8/3) \pi (\eta / \bsl^2) C^E
\vect{E}$] gives the net mechanical-contact force acting {\em on} the
inclusion,
\begin{equation}
  \vect{f}^{m,E} = (8/3) \pi Z^E \ymod \vect{E} - 4 \pi (\eta /
  \bsl^2) C^E \vect{E}.
\end{equation}

\end{document}